\documentclass[]{article}% insert '[draft]' option to show overfull boxes

\usepackage{varioref}%  smart page, figure, table, and equation referencing
\usepackage{wrapfig}%   wrap figures/tables in text (i.e., Di Vinci style)
\usepackage{threeparttable}% tables with footnotes
\usepackage{dcolumn}%   decimal-aligned tabular math columns
\newcolumntype{d}{D{.}{.}{-1}}
\usepackage{nomencl}%   nomenclature generation via makeindex
\usepackage{subfigure}% subcaptions for subfigures
\usepackage{subfigmat}% matrices of similar subfigures, aka small mulitples
\usepackage{fancyvrb}%  extended verbatim environments
\fvset{fontsize=\footnotesize,xleftmargin=2em}
\usepackage{lettrine}%  dropped capital letter at beginning of paragraph
\usepackage[colorlinks]{hyperref}%  hyperlinks [must be loaded after dropping]
 \usepackage{url}

\usepackage{amssymb}           % add ams symbols stuff
\usepackage{graphicx}          % add graphics
\usepackage{amsmath}           % assumes amsmath package installed
\usepackage{url}
\usepackage{flafter}           % Cause floats to appear after environment.
\usepackage{siunitx}           % Provides standard formatting of SI units.
\usepackage{cite}
\usepackage{epsfig}            % for postscript graphics files
\usepackage{epstopdf}
\usepackage{color}
\usepackage{cancel}
\usepackage{mathrsfs}
\usepackage{pdflscape}
\usepackage{textcomp}
\usepackage[T1]{fontenc}
\usepackage[latin9]{inputenc}
\usepackage{geometry}
\usepackage{setspace}
\usepackage{float}
\usepackage{enumerate}

\usepackage{multicol} 

\usepackage{multirow}% http://ctan.org/pkg/multirow
\usepackage{hhline}% http://ctan.org/pkg/hhline 
\usepackage{booktabs}  
\usepackage[table,x11names]{xcolor}
\usepackage{fancyvrb}

\usepackage{siunitx}           % Provides standard formatting of SI units.

\newcommand\scalemath[2]{\scalebox{#1}{\mbox{\ensuremath{\displaystyle #2}}}} %
\usepackage{setspace}
\usepackage{etoolbox}

\BeforeBeginEnvironment{equation}{\begin{singlespace}}
	\AfterEndEnvironment{equation}{\end{singlespace}\noindent\ignorespaces}
\BeforeBeginEnvironment{align}{\begin{singlespace}}
	\AfterEndEnvironment{align}{\end{singlespace}\noindent\ignorespaces}

\usepackage{tikz}
\usetikzlibrary{shapes.geometric}
\usetikzlibrary{shapes.arrows}

\usepackage{hyperref}
\usepackage[hypcap]{caption}
\hypersetup{colorlinks=true,linkcolor=blue,citecolor=blue,filecolor=blue,urlcolor=black}

\title{Rudder Augmented Trajectory Correction for Small UAV to Minimize Lateral Image Errors }
\author{
	Thomas Fisher\thanks{Graduate Student, Mechanical and Aerospace Engineering, Utah State University, Logan, UT, 84332,  USA}
	\and
	Rajnikant Sharma\thanks{Assistant Professor, Aerospace Engineering, University of Cincinnati, Cincinnati,  OH, USA. email: rajnikant.sharma@uc.edu}}

\newcommand{\nl}{\vspace{5mm} \\}

 \newcommand{\nlss}{\vspace{15mm}}

\definecolor{Gray}{gray}{0.95}

\begin{document}
	
\maketitle
	%\nodate
	%\nodate{}
	%

	\begin{abstract}
		%Put abstract here if you need one. 
		
		Civil applications for unmanned aerial vehicles (UAVs) have increased rapidly over the last few years.   In the realm of civil applications, many aircraft carry cameras that are physically fixed to the airframe.     While this yields a simple and robust remote sensing platform, the imagery quality diminishes with increasing attitude errors. A rudder augmented trajectory correction method for small unmanned aerial vehicles is discussed in this paper.   The goal of this type of controller is to minimize the lateral image errors of body fixed non-gimbaled  cameras.   We present a comparison to current aileron only trajectory correction autopilots.   Simulation and flight test results are presented that show significant reduction in the roll angle present during trajectory correction resulting in a large effect on total  flight line image deviations.

	%	Civil applications for unmanned aerial vehicles (UAVs) have increased rapidly over the last few years.   In the realm of civil applications, many aircraft carry cameras that are physically fixed to the airframe.     While this yields a simple and robust remote sensing platform, the imagery quality diminishes with increasing attitude errors.   This paper addresses both lateral (flight line) and the roll angle ($\phi$) errors of the aircraft orientation.  In the past, most autopilot design has focused on minimizing trajectory errors through the use of aileron control surfaces only, banking the wings in the direction of desired course.   While there are many benefits to this, for applications that are very sensitive to attitude deviations, rudder augmented trajectory control (RATC) offers significant reductions in imagery error by reducing the potentially large bank angle ($\phi$) required in aileron only control schemes.  In conclusion, partial or full implementation of RATC should not be overlooked in UAV trajectory controller implementation when the UAV application is sensitive attitude deviations.		
	\end{abstract}

	\section*{Nomenclature}
	
	\begin{tabbing}
		XXXXXXXXX \= \kill% this line sets tab stop
		$\alpha$ \>  Aircraft angle of attack (AOA), $radians$\\
		$\beta$ \>  Aircraft sideslip angle, $radians$ \\
		$\rho$ \> Air density, $N/m^2$\\
		$S_w$ \> Surface area of the Aircraft wing, $m^2$\\
		$b_w$ \> Reference wing chord length, $m$\\
		$p_n$ \> Location of the aircraft relative to the North Earth fixed coordinate axis, $m$\\
		$p_e$ \> Location of the aircraft relative to the East Earth fixed coordinate axis, $m$\\
		$p_d$ \> Location of the aircraft relative to the Down Earth fixed coordinate axis, $m$\\
		$x$ \> Body coordinate axis pointing out the nose of aircraft\\
		$y$ \> Body coordinate axis pointing out the right wing of aircraft\\
		$z$ \> Body coordinate axis pointing out the bottom of aircraft\\
		$u$ \> Aircraft  linear velocity in the $x$ axis direction relative to surrounding air, $m/s$\\
		$v$ \> Aircraft linear velocity in the $y$ axis direction relative to surrounding air, $m/s$ \\
		$w$ \> Aircraft linear velocity in the $z$ axis direction relative to surrounding air, $m/s$ \\
		$p$ \> Aircraft angular velocity about  $x$ axis (rolling rate), $radians/s$\\
		$q$ \> Aircraft angular velocity about  $y$ axis (pitching rate), $radians/s$\\
		$r$ \> Aircraft angular velocity about  $z$ axis (yawing rate), $radians/s$\\
		$l$ \> Aircraft rolling moment, $ N/m$\\
		$m$ \> Aircraft pitching moment,  $N/m$\\
		$n$ \> Aircraft yawing moment, $N/m$\\
		$F_x$ \> Aerodynamic force in the $x$ axis direction (including thrust), $N$\\
		$F_y$ \> Aerodynamic force in the $y$ axis direction (including thrust), $N$\\
		$F_z$ \> Aerodynamic force in the $z$ axis direction (including thrust), $N$\\	
			$M_x$ \> Aerodynamic moment in the $x$ axis direction (including thrust), $N$\\
			$M_y$ \> Aerodynamic moment in the $y$ axis direction (including thrust), $N$\\
			$M_z$ \> Aerodynamic moment in the $z$ axis direction (including thrust), $N$\\		
		$\phi$ \> Euler angle of rotation about $x$ axis (bank angle), $radians$\\
		$\theta$ \> Euler angle of rotation about $y$ axis (pitch angle), $radians$\\
		$\psi$ \> Euler angle of rotation about $z$ axis (yaw angle), $radians$\\
		$\chi$ \> Trajectory course angle, $radian$s \\
		$R$ \> Turning Radius, $m$\\
		$g$ \> Gravity, $m/s^2$\\
		$V_a$\> Aircraft total velocity relative to surrounding air, $m/s$\\
		$V_g$\> Aircraft total velocity relative to the ground, $m/s$\\
		$W_n$ \> Wind in North Earth fixed coordinate axis, $m/s$\\
		$W_e$ \> Wind in East Earth fixed coordinate axis, $m/s$\\
		$W_d$ \> Wind in Down Earth fixed coordinate axis, $m/s$\\
		$\gamma$ \> Angle between the aircraft trajectory and the horizon (climb angle), $ radians$\\
		$Error_{total}$ \> Total image error from the desired flight line, $ m$\\
		$Error_{lateral}$ \> Image error due to lateral trajectory errors, $m$\\
		$\bar{h}$ \> Aircraft altitude reference from the local ground level, $ m$\\
		$R_{interial}^{Path}$ \> Rotation matrix from the desired path line to the inertial frame\\
		$e_{px}$ \> Error from a desired flight line with respect to $x$ , $m$\\
		$e_{py}$ \> Error from a desired flight line with respect to $y$ (lateral trajectory error), $m$\\
		$e_{pz}$ \> Error from a desired flight line with respect to $z$, $m$\\
		$p^i$ \> Vector pointing to the current aircraft position with respect to Earth center\\
		$r^i$ \> Vector pointing to the desired flight line with respect to Earth center\\
		$e_{orbit}$ \> Lateral error from a desired flight orbit, $m$\\
		$c_n$  \> Location of the desired orbit with respect to the North Earth fixed coordinate axis, $ m$\\
		$c_e$  \> Location of the desired orbit with respect to the East Earth fixed coordinate axis, $m$\\
		$\lambda$ \> Orbit direction (+1 clockwise and -1 counterclockwise)\\
		$r_d$ \> Commanded orbit radius, $m$\\
		$I_{xx}$ \> Moment of inertia, $kg\ m^2$\\
		$I_{yy}$ \> Moment of inertia, $kg\ m^2$\\
		$I_{zz}$ \> Moment of inertia, $kg\ m^2$\\
		$I_{xy}$ \> Product of inertia, $kg\ m^2$\\
		$I_{xz}$ \> Product of inertia, $kg\ m^2$\\
		$I_{yz}$ \> Product of inertia, $kg\ m^2$\\
		$C_{l_{0}}$ \> Inherent rolling moment coefficient \\ 
		$C_{l_{\beta}}$ \> Change in $l$ with respect to $\beta$\\
		$C_{l_{p}}$ \> Change in $l$ with respect to $p$\\
		$C_{l_{r}}$ \> Change in $l$ with respect to $r$\\
		$C_{l_{\delta_a}}$ \>Change in $l$ with respect to $\delta_a$\\
		$C_{l_{\delta_r}}$ \> Change in $l$ with respect to $\delta_r$\\
		$C_{n_{0}}$ \> Inherent yawing moment coefficient \\ 
		$C_{n_{\beta}}$ \> Change in $n$ with respect to $\beta$\\
		$C_{n_{p}}$ \> Change in $n$ with respect to $p$\\
		$C_{n_{r}}$ \> Change in $n$ with respect to $r$\\
		$C_{n_{\delta_a}}$ \>Change in $n$ with respect to $\delta_a$\\
		$C_{n_{\delta_r}}$ \> Change in $n$ with respect to $\delta_r$\\
		$C_{Y_{\beta}}$ \> Change in aerodynamic side force with respect to $\beta$

	\end{tabbing}

\section{Introduction}

The use of small unmanned aerial vehicles (UAVs) in remote sensing programs is growing tremendously \cite{Al-Arab2013}.    Often, these UAVs carry small scientific cameras to gather aerial imagery\cite{Jensen2011} \cite{Mckee2011}.	
This imagery is used in a variety of civil and environmental engineering applications \cite{barfuss2012evaluation}. Focusing on situations specifically dealing with aerial surveying, payload performance is of utmost importance since multiple images have to be stitched together into large  single images or ``mosaics.''   While this is trivial with a small number of images, UAV image datasets can be very large (often 500+ images).   To decrease the time required to create the final mosaic, accurate position and correct orientation of the payload during flight is necessary.  This brings to light two critical aspects  of UAV performance.   First, the ability to follow the desired flight plan and second, keeping the camera pointed in the right direction.   This paper focuses on the specific case of aerial surveying applications that require the camera sensor to be as parallel with the ground as possible.  With that in mind, payload performance is  maximized to the degree in which the UAV keeps the camera level and positioned over the region of interest (ROI).  
%\subsection{Trajectory Correction and the Impact on Payload Imagery}

In the past,  aileron only trajectory correction (AOTC) schemes have been successfully implemented in small UAVs \cite{Beard2012} \cite{Reed2004}.  They are simple, robust, and very well understood from both human-piloted and UAV perspectives.  Since aircraft are generally designed to be stable in roll \cite{Phillips2009}, AOTC methods are very robust.  The primary means of correcting trajectory errors is by banking the wing to one side and allowing the lateral component of the lift vector to provide the restoring side force leading to a lateral acceleration back to the desired flight line.  These bank angles can be greater than $\phi > 45^{\circ}$.  The higher the bank angle, the larger the restoring side force and the better trajectory correction performance of the AOTC method, reducing cross track lateral image error.   However, this puts AOTC in direct conflict with payload performance since the use of wing bank angle to correct trajectory deviations introduces another source of lateral image error. As the aircraft rolls ($\phi \ne 0$), the center of the image is no longer aligned with the desired image location.  In some cases, the bank angle can be extreme enough that the desired ground image is no longer in the camera  field of view.  This lateral shift in the imagery is aggravated as the above ground level (AGL) altitude increases.   The roll angle creates considerable lateral image error with a seemingly small amount of bank angle (see Figure \ref{fig:FOV1}).  Since this trajectory correction method inherently introduces lateral error, two general solutions have historically been implemented to correct this roll distortion in the aerial imagery.  

The first is the use of mechanical stabilization or ``gimballing'' of the camera \cite{Semke2008}.   Gimbals work by rotating the camera independent of the aircraft body-fixed  coordinate frame. They can be adapted to correct only bank ($\phi$) deviations or can be configured to correct both pitch and roll changes. While they can yield significant reductions in image error, they add additional complexity and weight to the UAV payload.   As the ratio of UAV  maximum takeoff weight (MTOW) to camera/payload weight increases, this leaves little room for addition mechanical stabilization.   In addition, to accurately geo-reference the aerial imagery, they require a separate inertial measurement unit (IMU) to determine the orientation of the camera focal plane.

The second method is to use software to correct the shifted/rotated imagery \cite{Jensen2009} \cite{Wang2011}.    In a general sense, this works by mapping the deviated image to a flat plane equivalent.  However, if the end result is scientific-grade data, stretching and compressing of the image pixels while mapping to an equivalent image can cause data loss.  Such software correction can be very computationally intensive.   This impacts applications dealing with real-time data correction and increases the time needed to generate mosaics. 

In this paper, instead of trying to work around the issues inherent in AOTC schemes, we use a different trajectory correction algorithm designated as Rudder Augmented Trajectory Correction (RATC).  In effect, it eliminated the root cause of the lateral error due to roll angle found in AOTC.    The focus is on low weight non-gimballed (``fixed'') camera payloads and the use of the rudder control surface to provide the trajectory correction.  

%\subsection{Rudder Augmented Trajectory Correction: Overview}	
Rudder augmented trajectory correction (RATC) works by using the ailerons to maintain wings level flight.   As the UAV drifts off the desired course, instead of banking the aircraft with the ailerons, the rudder surface is used to apply the correction. Since the vertical stabilizer is located behind the center of gravity, this produces a yawing moment.  As a result, the sideslip angle $\beta$ (airflow relative to the aircraft axial plane) becomes nonzero. The fuselage side force vector,  $C_{Y_{\beta}}$, then also becomes nonzero and provides the restoring side force.  Using Newtons second law, this side force results in lateral acceleration back to the desired flight line.   The general principles behind RATC are not new concepts. Nearly all full scale aircraft use rudder trim tabs for long range path trim.   As early as 1935 the German firm Siemens implemented a rudder course control unit that was tied directly to a magnetic compass.  While the control unit was effective at holding course over long distances, little mention was given to the amount of control that these systems had and their ability for tight trajectory correction  \cite{Duane2014}.  In more recent research, it has been shown that the rudder control surface can correct heading deviations faster than its AOTC counterpart \cite{Ahsan2012}.  This effect comes from two fundamental differences between AOTC and RATC control schemes.  First, the rudder input applies a direct moment on the aircraft along the z-axis creating a change in heading.   Second, as shown later, RATC does not require successive loop closure.  

This paper expands on the idea of using the rudder as the primary means of correcting course deviations and presents a control strategy that allows for aggressive trajectory following. Initial simulation results of RATC were presented in AIAA GNC 2016~\cite{fisher2016rudder}. 	In this paper,  a comparison to current aileron only trajectory correction autopilots and the analytical derivation of the rudder augmented trajectory correction controller is presented.    Effects of rudder control input on roll/bank angle is also investigated.   Simulation and flight test results are presented that show significant reduction in the roll angle present during trajectory correction.

The rest of the paper  is organized as follows: In Section~\ref{s:Problem Formulation} the problem is formulated and equations of motion UAV and path and image errors are included. The AOTC controller is introduced in Section~\ref{s: AOTC}.  The RATC controller is derived in Section~\ref{s: RATC}. Flight tests and results are presented in Section~\ref{s: Flight Tests}.  Section~\ref{s: Flight Paths} discusses implications of RATC on path planning algorithms. Finally, we present our conclusions in Section~\ref{s: Conclusions}.

\section{Problem Formulation} \label{s:Problem Formulation}
The generalized nonlinear 6-degree of freedom (DOF) equations of motion for an aircraft are shown in  \eqref{eq:gen1}-\eqref{eq:gen4}.  For further explanation and derivation see \cite{Phillips2009}.   %The sign conventions for the 12 states of \eqref{eq:gen1}-\eqref{eq:gen4} as well as the control surface deflections can be found in Figures \ref{fig:coordinate} and \ref{fig:deflection}.

%	\subsection{Aircraft Mathematical Model}  
Equation \eqref{eq:gen1} shows the linear acceleration of the aircraft with respect to the body axis.	
\begin{equation}
\label{eq:gen1}
\frac{W}{g}
\begin{Bmatrix}
\dot{u}\\
\dot{v}\\
\dot{w}\\
\end{Bmatrix}=
\begin{Bmatrix}
F_{x}\\
F_{y}\\
F_{z}\\
\end{Bmatrix}+
\frac{W}{g}
\begin{Bmatrix}
rv-qw\\
pw-ru\\
qu-pv\\
\end{Bmatrix}
\end{equation}

Equation \eqref{eq:gen2} shows the angular acceleration of the aircraft with respect to the body axis.
\begin{eqnarray}
\scalemath{0.91}{
	\label{eq:gen2}
	\begin{array}{*{20}{l}}
	\begin{bmatrix}
	I_{xx}&-I_{xy}&-I_{xz}\\
	-I_{xy}& I_{yy}&-I_{yz}\\
	-I_{xz}&-I_{yz}& I_{zz}\\
	\end{bmatrix}			
	\begin{Bmatrix}
	\dot{p}\\
	\dot{q}\\
	\dot{r}\\
	\end{Bmatrix}=
	\begin{Bmatrix}
	M_{x}\\
	M_{y}\\
	M_{z}\\
	\end{Bmatrix}+	
	%				+\begin{bmatrix}
	%				0&-h_{z_{b}}&h_{y_{b}}\\
	%				h_{z_{b}}& 0&-h_{x_{b}}\\
	%				-h_{y_{b}}&h_{x_{b}}&0\\
	%				\end{bmatrix}
	%				\begin{Bmatrix}
	%				p\\
	%				q\\
	%				r\\
	%				\end{Bmatrix}
	\begin{bmatrix}
	\begin{pmatrix}
	I_{yy}-I_{zz}
	\end{pmatrix} qr + I_{yz}
	\begin{pmatrix}
	q^{2}-r^{2}\\
	\end{pmatrix}
	+I_{xz}pq-I_{xy}pr\\		
	\begin{pmatrix}
	I_{zz}-I_{xx}
	\end{pmatrix} pr + I_{xz}
	\begin{pmatrix}
	r^{2}-p^{2}\\
	\end{pmatrix}
	+I_{xy}pr-I_{yz}pq\\		
	\begin{pmatrix}
	I_{xx}-I_{yy}
	\end{pmatrix} pq + I_{xy}
	\begin{pmatrix}
	p^{2}-q^{2}\\
	\end{pmatrix}
	+I_{yz}pr-I_{xz}qr\\
	\end{bmatrix}
	\end{array}
}
\end{eqnarray}
Equation \eqref{eq:gen3} shows the velocity of the aircraft with respect to the Earth, including wind perturbations, $W$.
\begin{eqnarray}
\label{eq:gen3}
\begin{array}{*{20}{l}}
\begin{Bmatrix}
\dot{p_n}\\
\dot{p_e}\\
\dot{p_d}\\
\end{Bmatrix}=		
\begin{Bmatrix}
W_n\\
W_e\\
W_d\\
\end{Bmatrix}+		
\begin{bmatrix}
C_{\theta}C_{\psi}&S_{\phi}S_{\theta}C_{\psi}-C_{\theta}S_{\psi}&C_{\phi}S_{\theta}C_{\psi}+S_{\phi}S_{\psi}\\
C_{\theta}S_{\psi}&S_{\phi}S_{\theta}S_{\psi}-C_{\phi}C_{\psi}&C_{\phi}S_{\theta}S_{\psi}-S_{\phi}S_{\psi}\\
C_{\theta}C_{\psi}&S_{\phi}S_{\theta}C_{\psi}-C_{\theta}S_{\psi}&C_{\phi}S_{\theta}C_{\psi}+S_{\phi}S_{\psi}\\
\end{bmatrix}
\begin{Bmatrix}
u\\
v\\
w\\
\end{Bmatrix}		
\end{array}
\end{eqnarray}
Equation \eqref{eq:gen4} shows angular velocity of the aircraft Euler angles.
\begin{equation}
\label{eq:gen4}
\begin{Bmatrix}
\dot{\phi}\\
\dot{\theta}\\
\dot{\psi}\\
\end{Bmatrix}=
\begin{bmatrix}
1&S_{\phi}S_{\theta}/C_{\theta}&C_{\phi}S_{\theta}/C_{\theta}\\			
0&C_{\phi}&-S_{\phi}\\		
0&S_{\phi}/C_{\theta}&C_{\phi}/C_{\theta}\\
\end{bmatrix}
\begin{Bmatrix}
p\\
q\\
r\\
\end{Bmatrix}
\end{equation}
\subsection{Image Errors}
\begin{figure}[h]
	%	\centering
	\includegraphics[width=14cm]{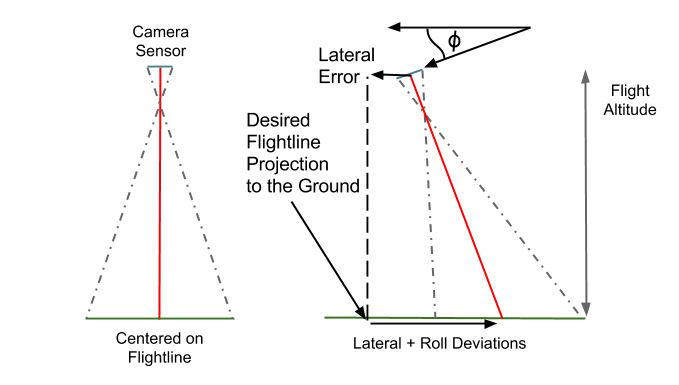}
	\caption{Imagery Errors from Lateral and Roll Deviations}
	\label{fig:FOV1}
\end{figure}	

The image error \eqref{eq:FOVerror1} of the fixed camera imagery is modeled as two separate components as shown in Figure \ref{fig:FOV1}.  
\begin{equation}
\label{eq:FOVerror1}
Error_{total} = Error_{lateral}+ \bar{h} \  tan(\phi)
\end{equation}
The first component,  $Error_{lateral}$,  is the lateral error from the desired flight line.  It can also be described as the error in the ground track if viewed from above the aircraft.   The second component, the image error due to bank angle ($\phi$), is derived from the geometry of the aircraft attitude.   The significance of this second term is directly related to the aircraft Above Ground Level (AGL) altitude.    As the altitude increases, this second term can quickly become the dominant error source. Since the aircraft is operating in 3-dimensional space, equation \eqref{eq:FOVerror1} can be defined in vector notation.     However, in aerial imaging, constant distance from the ground is preferred to keep the Ground Spatial Distance (GSD) the same.  As a result, this thesis only deals with the portions of the flight where the altitude is held steady and the pitch angle is negligible ($\theta \approx 0$).  For the purposes of this paper, \eqref{eq:FOVerror1} will only refer to scalar components of the lateral and roll image error and neglect pitch and yaw error contributions.

\subsection{Path Errors}
The path error,  $Error_{lateral}$,   can be mathematically modeled with regards to both straight line and orbit following flight. It is  important to note that the two error sources from \eqref{eq:laterror1}  never occur at the same time since they are functions of the path planning algorithms in the autopilot.
\begin{equation}
\label{eq:laterror1}
Error_{lateral} = 	e_{py}+e_{orbit} 
\end{equation}

\subsubsection{Straight Line Flight}

For straight line flight, the lateral error, $e_{py}$, can be described as the distance perpendicular to the desired flight path \cite[p.~175-176]{Beard2012}.   The vectors $p^i$ and $r^i$ in \eqref{eq:pathline} should not be confused with the rolling and yawing rates of the aircraft.  The vector $p^i$ points to the current location of the UAV and  $r^i$ is the vector pointing to the start of the desired flight line.  
\begin{equation}
\label{eq:pathline}
e_{py} \rightarrow
\begin{bmatrix}
e_{px}\\
e_{py}\\
e_{pz}\\
\end{bmatrix}= R_{inertial}^{Path}(p^i-r^i)
\end{equation}

\subsubsection{Orbiting or Circular Flight}
The lateral error  during orbit or circular trajectory following ($e_{orbit}$) can be described as the difference in the commanded radius versus the actual circular path of the UAV \eqref{eq:pathorbit1} where,  $\sqrt{(p_n-c_n)^2+(p_e-c_e)^2}$ is the distance or "radius" of the UAV from the center of the commanded orbit.  Using a orbit direction factor $\lambda $  ($\lambda $ =1  $\rightarrow$ CW orbit, $\lambda $ = -1 $\rightarrow$ CCW orbit),  error is defined as positive out the right wing of the aircraft regardless of the orbit direction \cite[p.181]{Beard2012}.

\begin{equation}
\label{eq:pathorbit1}
e_{orbit} = -\lambda
\renewcommand\arraystretch{1.2}
\begin{pmatrix}
r_d-\sqrt{(p_n-c_n)^2+(p_e-c_e)^2} 
\end{pmatrix}
\end{equation}

\section{AOTC Controller}\label{s: AOTC}
To help illustrate the differences between AOTC and RATC control schemes, a brief introduction to the equations of motion governing AOTC flight will be presented. AOTC is based on the aerodynamic principle that for coordinated turns ($\beta = 0$), a relationship exists between the bank angle ($\phi$), airspeed ($V_a$), gravity ($g$), the climb angle ($\gamma$), and the turning radius $R$ of the aircraft \cite[p.~281-283]{Phillips2009}.   This relationship is defined in Equation \eqref{eq:coordTurn}.  
\begin{equation}
\label{eq:coordTurn}
R = \frac{V_a^2 cos\gamma}{g \ tan\phi}
\end{equation}
For level flight trajectory correction we assume that $\gamma \approx 0$ and using small angle approximations this yields: 
\begin{equation}
\label{eq:coordTurn2}
R = \frac{V_a^2}{g \ tan\phi}
\end{equation}
Assuming no wind, $V_a = V_g$ (see \cite[p.23]{Beard2012}), the heading rate of change $\dot{\psi}$  can be described with respect to the desired circular radius and the aircraft velocity \eqref{eq:psidot}.   During the coordinated turn ($\beta = 0$), the change rate in aircraft heading is equivalent to the change rate in trajectory course $\dot{\psi} =\dot{\chi}$.
\begin{equation}
\label{eq:psidot}
\dot{\chi} = \dot{\psi} = \frac{V_a}{R} = \frac{V_g}{R}
\end{equation}
%$\frac{V_a}{R} $ can be interpreted geometrically as the angular measurement of a circle divided by the time which it would take to traverse the complete circle described by $R$ at a speed $V$.  This yields an angular rate of time measurement.  It can also be understood in terms of cross products.\\\\		
Substituting \eqref{eq:coordTurn2} into \eqref{eq:psidot} yields:
\begin{equation}
\label{eq:coordTurn3}
\dot{\chi} = \frac{g \ tan\phi}{V_g}
\end{equation}	

Using Laplace transforms and small angle approximations of $tan \phi$, a transfer function between the course and the roll angle is derived:
\begin{equation}
\label{eq:coordTurn4}
\frac{\chi }{\phi}	= \frac{g}{V_g s}
\end{equation}\\
Using the aerodynamic coefficients, transfer functions between $\phi$ and $\delta_{aileron}$ can also be developed.   At this point, successive loop closure can be applied to the outer (course) and inner (control surface) control loops defining the Aileron Only Trajectory Correction scheme (AOTC).  Since the inner loop must go to unity for successive loop closure to be valid, the outer loop is considerably slower.    While this doesn't introduce large amounts of lag in the natural frequency of the system, RATC doesn't require successive loop closure and has slightly faster response to heading correction \cite{Ahsan2012}.

\section{RATC Controller}\label{s: RATC}
The goal of trajectory control is to bring $\chi_{actual} = \chi_{desired}$. Expanding $\chi_{actual} $ and assuming no wind, $V_a = V_g$,   $\chi_{actual} = \psi+\beta$.   Treating $\beta$ as an input disturbance, RATC approximates  $\chi = \psi$.  The control design is simplified since there is only a single transfer function between the desired heading input  and the  control surface deflection.   This  eliminates the need for successive loop closure and  allows RATC to respond faster than its AOTC counterpart.  

\begin{figure}[H]
	\centering
	\includegraphics[width=8cm]{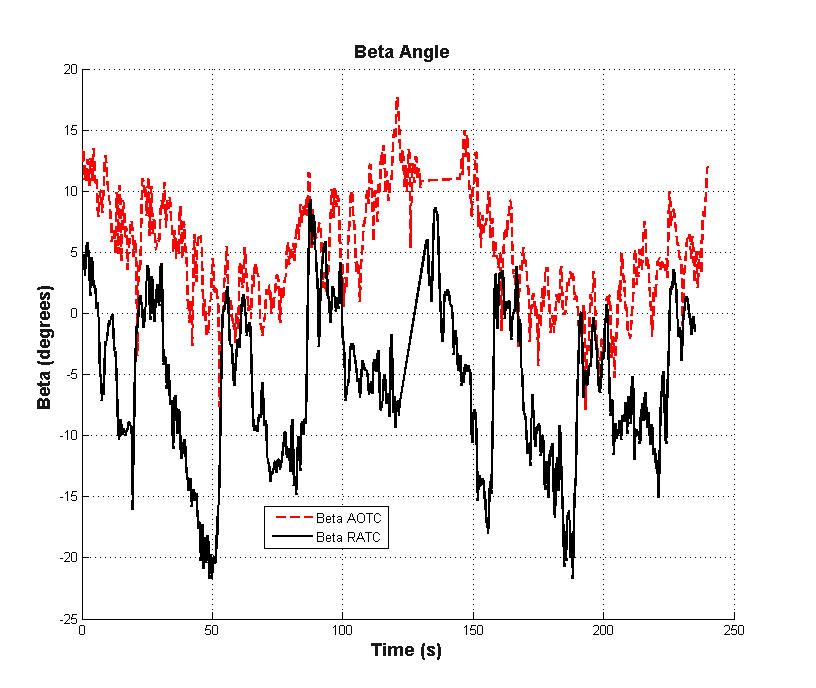}
	\caption{Beta Flight \#3}
	\label{fig:FR_beta1}
\end{figure}

Interested readers should be aware that RATC schemes are more sensitive to changes in the governing aerodynamic coefficients of the lateral force terms $C_Y$ than AOTC.   Depending on the aerodynamic configuration, these coefficients can cause significant deviations in controller performance when compared with the results shown in this thesis. The largest contributors are the side force generation terms due to sideslip ($\beta$).   If these are small, $\beta$ angles are higher, weakening the assumption made during the derivation of the controller that $\beta \approx 0$.  For example, the Aerosonde UAV used in the simulation had high side force generation terms and the corresponding $\beta$ angles were low (see Figure \ref{fig:plot4a}).   The Minion UAV used for flight testing had considerably smaller $C_Y$ terms and, as a consequence, the $\beta$ angles were $200-400$\% higher (see Figures  \ref{fig:FR_betaC} and \ref{fig:FR_beta1}).       Also, not all regimes of flight are suitable for RATC.   For instance, during climbing flight or when the assumptions that $q = 0$ \& $\dot{q} = 0$ are not valid, RATC can cause controller instabilities.  In short, the RATC algorithm as presented here is not as robust as the corresponding AOTC equivalent.

Starting with \eqref{eq:gen2} and assuming a symmetric aircraft,  $I_{xy}$ and $I_{yz} = 0$.  The inertia tensor on the right hand side of \eqref{eq:gen2} can now be written as follows:
\begin{equation}
\label{eq:int1}
\begin{bmatrix}
I_{xx}& 0 &-I_{xz}\\
0 & I_{yy}& 0\\
-I_{xz}& 0 & I_{zz}\\
\end{bmatrix}
\end{equation}\\
Assuming that the propeller thrust vector contributes no moment, the following assumption can be made:
\begin{equation}
\label{eq:int2}
\begin{Bmatrix}
M_{x}\\
M_{y}\\
M_{z}\\
\end{Bmatrix} = 
\begin{Bmatrix}
l\\
m\\
n\\
\end{Bmatrix}
\end{equation}\\
Taking the inverse of \eqref{eq:int1} and multiplying on the left hand side of \eqref{eq:gen2} along with  algebraic manipulation  \cite[p.~34-36]{Beard2012} yields:
\begin{equation}
\label{eq:rot}
\begin{bmatrix}
\dot{p}\\
\dot{q}\\
\dot{r}\\
\end{bmatrix}=
\begin{bmatrix}
\Gamma_1pq-\Gamma_2qr+\Gamma_3l+\Gamma_4n\\
\Gamma_5pr-\Gamma_6(p^2-r^2)+\frac{m}{I_{yy}}\\
\Gamma_7pq-\Gamma_1qr+\Gamma_4l+\Gamma_8n\\
\end{bmatrix}	  
\end{equation}
Where the reduced moment of inertia terms \cite[p.36]{Beard2012} in \eqref{eq:rot} are defined as:
\begin{align*}
\label{eq:gen31a}
\Gamma_1 & = \frac{I_{xz}(I_{xx} - I_{yy}+ I_{zz})}{I_{xx}I_{zz} - I_{xz}^2}\nonumber \\ 
\Gamma_2 & = \frac{I_{zz}(I_{zz} - I_{yy})+ I_{xz}^2}{I_{xx}I_{zz} - I_{xz}^2}\nonumber \\
\Gamma_3 & = \frac{I_{zz}}{I_{xx}I_{zz} - I_{xz}^2}\nonumber \\
\Gamma_4 & = \frac{I_{xz}}{I_{xx}I_{zz} - I_{xz}^2}\nonumber \\
\Gamma_5 & = \frac{I_{zz} - I_{xx} }{I_{yy}}\nonumber \\
\Gamma_6 & = \frac{I_{xz} }{I_{yy}}\nonumber \\
\Gamma_7 & =  \frac{(I_{xx} - I_{yy})I_{xx}+ I_{xz}^2}{I_{xx}I_{zz} - I_{xz}^2}\nonumber\\
\Gamma_8 & =  \frac{I_{xx} }{I_{xx}I_{zz} - I_{xz}^2}\nonumber
\end{align*}
Focusing on the  $[\dot{r} =\Gamma_7pq-\Gamma_1qr+\Gamma_4l+\Gamma_8n] $ term from \eqref{eq:rot} we can further define $l$ and $n$ with a series of aerodynamic derivatives\cite[p.58]{Beard2012}:
\begin{equation}
\label{eq:r1}
l = \frac{1}{2} \rho V_a^2S_wb_w[C_{l_{0}}+C_{l_{\beta}}\beta+C_{l_{p}} \frac{b_wp}{2V_a}+C_{l_{r}} \frac{b_wr}{2V_a}+C_{l_{\delta_a}}\delta_a+C_{l_{\delta_r}}\delta_r]
\end{equation}
\begin{equation}
\label{eq:r2}
n = \frac{1}{2} \rho V_a^2S_wb_w[C_{n_{0}}+C_{n_{\beta}}\beta+C_{n_{p}} \frac{b_wp}{2V_a}+C_{n_{r}} \frac{b_wr}{2V_a}+C_{n_{\delta_a}}\delta_a+C_{n_{\delta_r}}\delta_r]
\end{equation}\\
Multiplying the inertia terms into \eqref{eq:r1} and \eqref{eq:r2} and combining like terms yields the following reduced equation:
\begin{equation}
\label{eq:r3}
\dot{r} = \Gamma_7pq-\Gamma_1qr +\frac{1}{2}\rho V_a^2S_wb_w [C_{r_0}+C_{r_\beta}\beta+C_{r_p}\frac{b_wp}{2V_a}+C_{r_r}\frac{b_wr}{2V_a}+C_{r_{\delta_a}}\delta_a+C_{r_{\delta_r}}\delta_r]
\end{equation}
Where the combined terms from \eqref{eq:r3} are defined as \cite[p.62]{Beard2012}:
\begin{align*}
\label{eq:red}
C_{r_0} & = \Gamma_4C_{l_0}+\Gamma_8C_{n_0}\nonumber \\ 
C_{r_\beta} & = \Gamma_4C_{l_\beta}+\Gamma_8C_{n_\beta}\nonumber \\ 
C_{r_p} & = \Gamma_4C_{l_p}+\Gamma_8C_{n_p}\nonumber \\ 
C_{r_r} & = \Gamma_4C_{l_r}+\Gamma_8C_{n_r}\nonumber \\ 
C_{r_{\delta_a}} & = \Gamma_4C_{l_{\delta_a}}+\Gamma_8C_{n_{\delta_a}}\nonumber \\ 
C_{r_{\delta_r}}& = \Gamma_4C_{l_{\delta_r}}+\Gamma_8C_{n_{\delta_r}}\nonumber  
\end{align*}
Using the last term of \eqref{eq:gen4}:
\begin{equation}
\label{eq:gen11}
\dot{\psi} = q\frac{sin(\phi)}{cos (\theta)}+ r \frac{cos(\phi)}{cos(\theta)}
\end{equation}\\
And applying small angle approximations about the trim conditions ($\phi$ and $\theta \approx 0$)  yields \eqref{eq:gen11a}:
\begin{equation}
\label{eq:gen11a}
\dot{\psi} = r+ q\phi
\end{equation}
Taking the derivative gives \eqref{eq:gen21}:
\begin{equation}
\label{eq:gen21}
\ddot{\psi} = \dot{r}+\dot{q}\phi+q\dot{\phi}
\end{equation}\\
Using trim conditions $q = 0$ and $\dot{q} = 0$, Equations  \eqref{eq:gen11a},  \eqref{eq:gen21}, and  \eqref{eq:r3} simplify to:
\begin{equation}
\label{eq:gen11a3}
\dot{\psi} = {r}
\end{equation}			
\begin{equation}
\label{eq:gen11a2}
\ddot{\psi} = \dot{r}
\end{equation}		
\begin{equation}
\label{eq:gen31}
\dot{r} = \frac{1}{2}\rho V_a^2S_wb_w [C_{r_0}+C_{r_\beta}\beta+C_{r_p}\frac{b_wp}{2V_a}+C_{r_r}\frac{b_wr}{2V_a}+C_{r_{\delta_a}}\delta_a+C_{r_{\delta_r}}\delta_r]
\end{equation}\\
Combining \eqref{eq:gen31} and  \eqref{eq:gen11a2} gives:	
\begin{equation}
\label{eq:gen41}
\ddot{\psi} = \frac{1}{2}\rho V_a^2S_wb_w [C_{r_0}+C_{r_\beta}\beta+C_{r_p}\frac{b_wp}{2V_a}+C_{r_r}\frac{b_wr}{2V_a}+C_{r_{\delta_a}}\delta_a+C_{r_{\delta_r}}\delta_r]
\end{equation}\\
Substituting \eqref{eq:gen11a3} into \eqref{eq:gen41} and rearranging yields:
\begin{equation}
\label{eq:gen51}
\ddot{\psi} = -a_{\psi_1}\dot{\psi}+a_{\psi_2}{\delta_r}+d_{\psi}
\end{equation}
where:
\begin{align*}
\label{eq:gen61}
a_{\psi_1}& = -\frac{\rho V_a^2S_wb_w^2C_{r_r}}{4} \nonumber\\
a_{\psi_2}& =\frac{1}{2}\rho V_a^2S_wb_w C_{r_{\delta_r}}\nonumber\\
d_{\psi}& = \frac{1}{2}\rho V_a^2S_wb_w [C_{r_0}+C_{r_\beta}\beta+C_{r_p}\frac{b_wp}{2V_a}+C_{r_{\delta_a}}\delta_a]\nonumber
\end{align*}		
In an effort to bring the nonlinear equations of motion into canonical $2^{nd}$ order plant model form, the disturbances packed inside the $  d_{\psi}$ term from \eqref{eq:gen11} are treated as input disturbances.  As a result, anytime $\beta$ is nonzero, slight trajectory tracking errors will be present.  Also, the propeller thrust vector does contribute to the side force terms and this contribution is directly related to $\beta$.   Since $\beta$ is assumed to be small, this effect is negligible.  However, as noted previously, if larger $\beta$ angles are present during trajectory tracking, this introduces another source of disturbance input not accounted for in the derivation of the RATC controller.  \nl
The Laplace transform of \eqref{eq:gen51} yields:
\begin{equation}
\label{eq:gen71}
{\psi} = \frac{a_{\psi_2}}{s(s+a_{\psi_1})}({\delta_r}+\frac{d_{\psi}(s)}{a_{\psi_2}})
\end{equation}\nlss\\
%		\pagebreak
%		Unstable pole at $0^-$\\ 
%		\begin{center}
%			\includegraphics[scale=0.4]{Pictures/siso_nocon}
%		\end{center}
Dropping the disturbance terms,  the transfer function mapping the rudder deflection directly to the aircraft heading is shown in \eqref{eq:gen71a}. 
\begin{equation}
\label{eq:gen71a}
\frac{\psi}{\delta_r} = \frac{a_{\psi_2}}{s(s+a_{\psi_1})}
\end{equation}\\
Applying a proportional derivative (PD) controller and enforcing the assumption that $\chi = \psi$ with $\beta \approx 0$ yields  \eqref{eq:gen81}. 
Since $\beta$ is often nonzero in lateral maneuvering, we will treat $\beta$ and the  force terms generated by $\beta$  as input disturbances shown in \eqref{eq:gen51}. 
\begin{equation}
\label{eq:gen81}
\frac{\chi}{\chi_{desired}} =
\frac{\psi}{\psi_{desired}} = \frac{k_{p_\psi}a_{\psi_2}}{s^2+s(a_{\psi_1}+a_{\psi_2}k_{d_\psi})+k_{p_\psi}a_{\psi_2}}
\end{equation}\\
Using $2^{nd}$ order design models and equating to \eqref{eq:gen81} results in a closed form solution for the controller proportional and derivative gains:
\begin{align}
\label{eq:gen61c}
k_{p_\psi}& = \frac{\omega_{n_\psi}^2}{a_{\psi_2}} \nonumber\\
k_{d_\psi}& = \frac{2\zeta_{\psi}\omega_{n_\psi}-a_{\psi_1}}{a_{\psi_2}}\nonumber\\
\end{align}

\section{Flight Test Results}\label{s: Flight Tests}
As a first step verification of the RATC controller, a simulation using a 6 degree of freedom aircraft mathematical model was implemented.   Matlab Simulink\textsuperscript{TM} environment was employed to numerically integrate the equations of motion (\ref{eq:gen1}- \ref{eq:gen4}).    An autopilot was also placed inside the simulation to allow control of the aircraft based on the methods found in \cite{Beard2012}.    The aircraft model used in the simulation was the Aerosonde UAV \cite{Beard2012}. Detailed results are included in~\cite{fisher2016rudder}. In this paper we just present the flight test results that comapres RATC and AOTC.

During data collection, a total of 22 flights were flown with 9.75 cumulative flight hours.   The flights began on 5/29/2015 and the final flight with the GoPro data collection was 11/23/2015.  The flights included preliminary rudder response testing, IMU upgrades, path manager changes, and trajectory testing with AOTC and RATC algorithms flown back to back.  
All the flights used  identical waypoints for both the AOTC and RATC tests.  With the RATC switch introduced in Section~\ref{s: RATC}, path planner would traverse an identical path for both types of controllers.   Using the same high level path management code for both algorithms ensured that the course command structure was consistent, and removed potential sources of bias when comparing controller performance.

The first flight dataset presented was flown on 10/28/2015 and was a large square figure eight.  Like in the simulation test cases, a fillet was placed in each corner to smooth the transition between flight lines.   The aircraft would start in the center of the crosspoint shown in Fig. \ref{fig:FR_position}.   It would complete the right hand square first by proceeding to fly to the bottom right-hand quadrant (south-east).   It would continue counter clockwise and cross back through the center flying towards the bottom left-hand quadrant (south-west).  After completing the left square in a clockwise fashion, it would complete the given flight plan at the center crosspoint.

The second flight plan shown was also flown on 10/28/2015 and was a large circle.   The purpose of this flight plan was twofold.   First, to demonstrate  RATC's ability to successfully fly a circular orbit without major instability issues.   Second, the motor current and velocity data was used to assess the RATC aerodynamic efficiencies and to provide recommendations regarding the use of the RATC algorithm as described in Section~\ref{s: RATC}.

The third flight plan shown was flown on 11/23/2015 and comprised a large rectangle with fillets in the corners.  While this flight seems to repeat flight plan \#1, two important distinctions  were made.  First, this flight plan was using a modified course planner algorithm that reduced the instability coming out of each corner. This change is very noticeable between Figure \ref{fig:FR_position} and Figure \ref{fig:FR_position1}.    Further explanation of this can be found in Section 8.6.   Second, this flight plan was flown with a GoPro HD camera providing a visual confirmation of the IMU data used to generate the graphs shown in this section.

\begin{figure}[H]
	\centering
	%\subfloat[]{%
	\includegraphics[width=5cm]{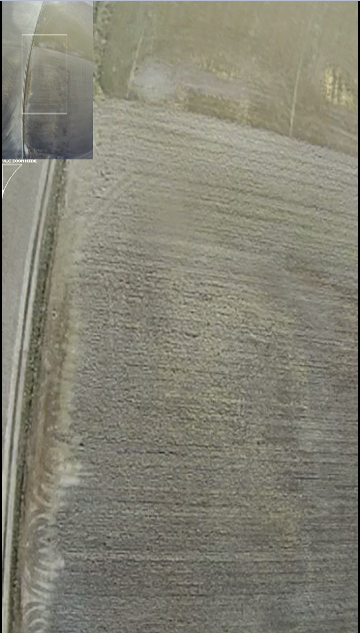}
	\includegraphics[width=5cm]{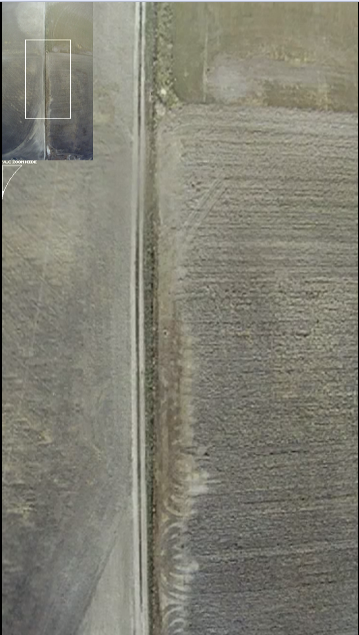}
	
	\caption{Comparison Image 1\\
		Left-AOTC, RATC-Right}
	
	\label{fig:comp1}
	\quad
\end{figure}

\begin{figure}[H]
	\centering
	%\subfloat[]{%
	\includegraphics[width=5cm]{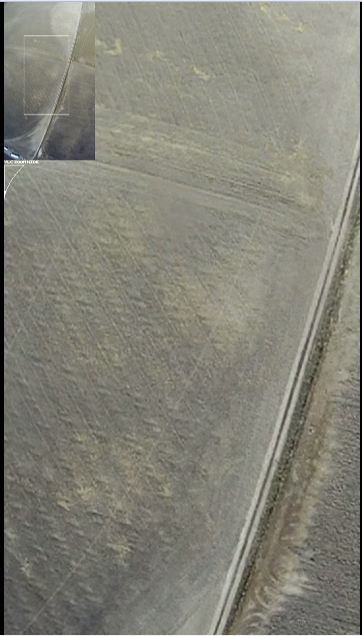}
	\includegraphics[width=5cm]{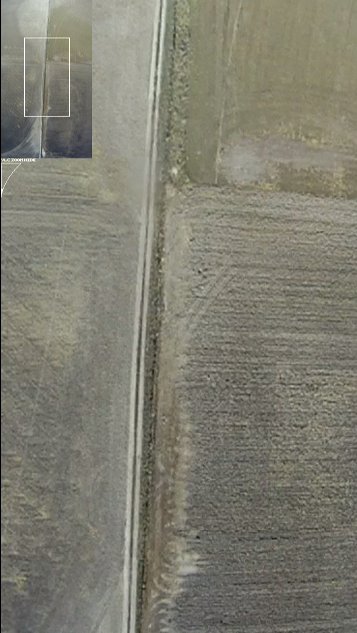}
	\label{fig:subfigure2}
	\caption{Comparison Image 2\\
		Left-AOTC, RATC-Right}
\end{figure}

\begin{figure}[H]
	\centering
	%\subfloat[First Image Set]{%
	\includegraphics[width=5cm]{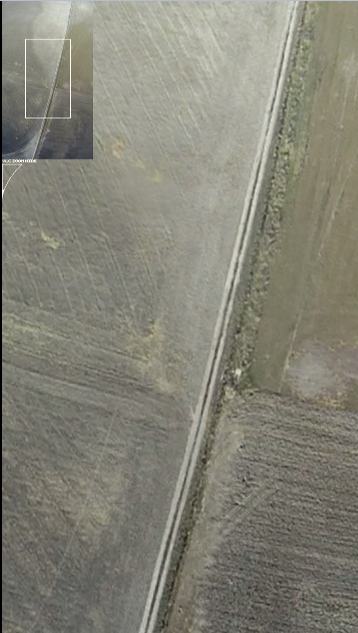}
	\includegraphics[width=5cm]{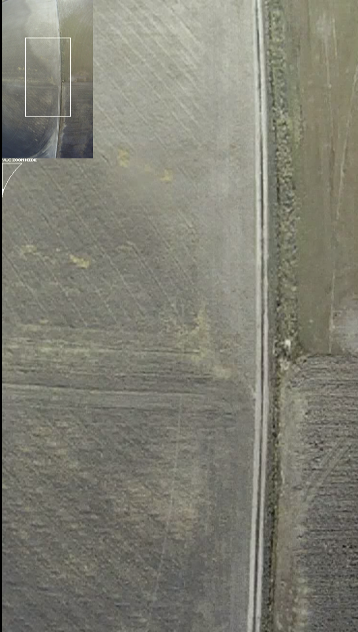}
	\label{fig:subfigure231}
	\caption{Comparison Image 3\\
		Left-AOTC, RATC-Right}
\end{figure}

\begin{figure}[H]
	\centering
	%	\subfloat[]{%
	\includegraphics[width=5cm]{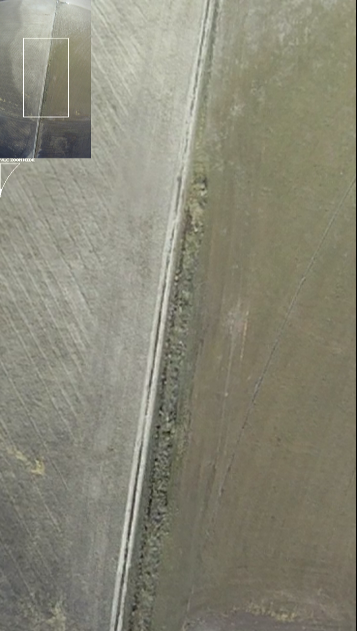}
	\includegraphics[width=5cm]{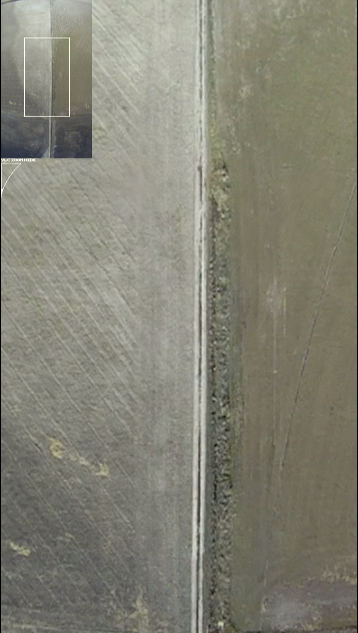}
	\label{fig:subfigure232}
	\caption{Comparison Image 4\\
		Left-AOTC, RATC-Right}
	%
	%	\caption{Main figure caption}
	%	\label{fig:figure23}
\end{figure}

\begin{figure}[H]
	\centering
	%\ContinuedFloat
	%	\subfloat[]{%
	\includegraphics[width=5cm]{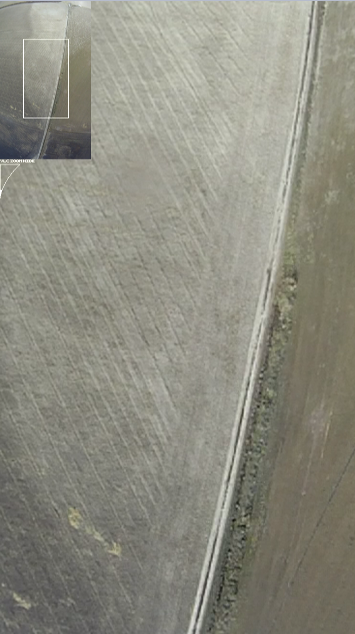}
	\includegraphics[width=5cm]{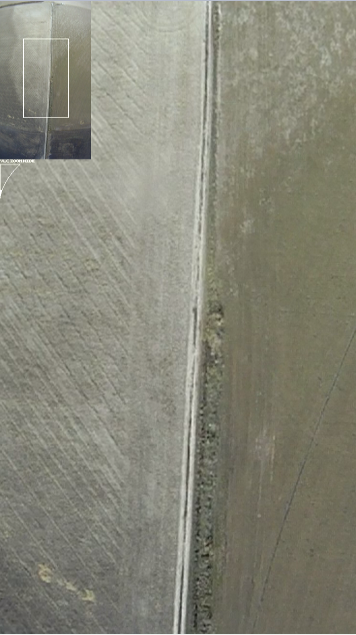}
	\label{fig:subfigure2312}
	\caption{Comparison Image 5\\
		Left-AOTC, RATC-Right}
\end{figure}

\begin{figure}[H]
	\centering
	%	\subfloat[]{%
	\includegraphics[width=5cm]{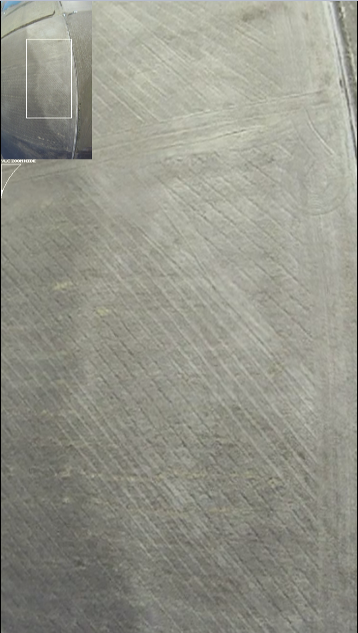}
	\includegraphics[width=5cm]{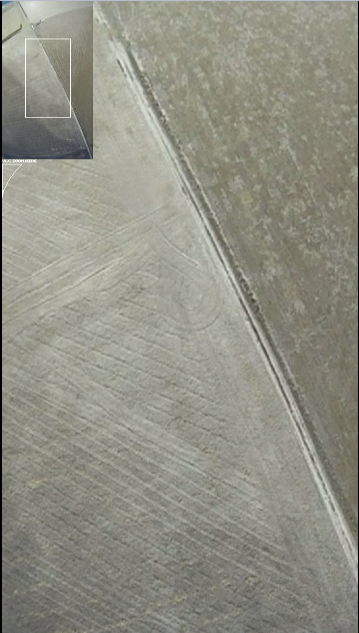}

	\caption{Comparison Image 6\\
		Left-AOTC, RATC-Right}
	\label{fig:comp6}

\end{figure}

As a final note, the flight data was collected at 150 m AGL but since  AggieAir\textsuperscript{TM} routinely flies at 450 m for aerial image collection, all of the image error plots for the 3 test cases are shown with the altitude incremented by 300 m to reflect this operational difference.  The GoPro flights were also conducted at lower altitudes for convenience, safety, and to ensure adequate ground resolution to see landmarks used as verification in the imagery.  For the purpose of comparing payload imagery in Figures \ref{fig:comp1} - \ref{fig:comp6}  to total error, Figure \ref{fig:imgErr_150_GP} was not incremented by the 300 m described previously.   To see the statistical results for both altitudes see Table \ref{tab:flight_results}.

\subsection{Flight Plan \#1}
\begin{figure}[H]
	\centering
	\includegraphics[width=8cm]{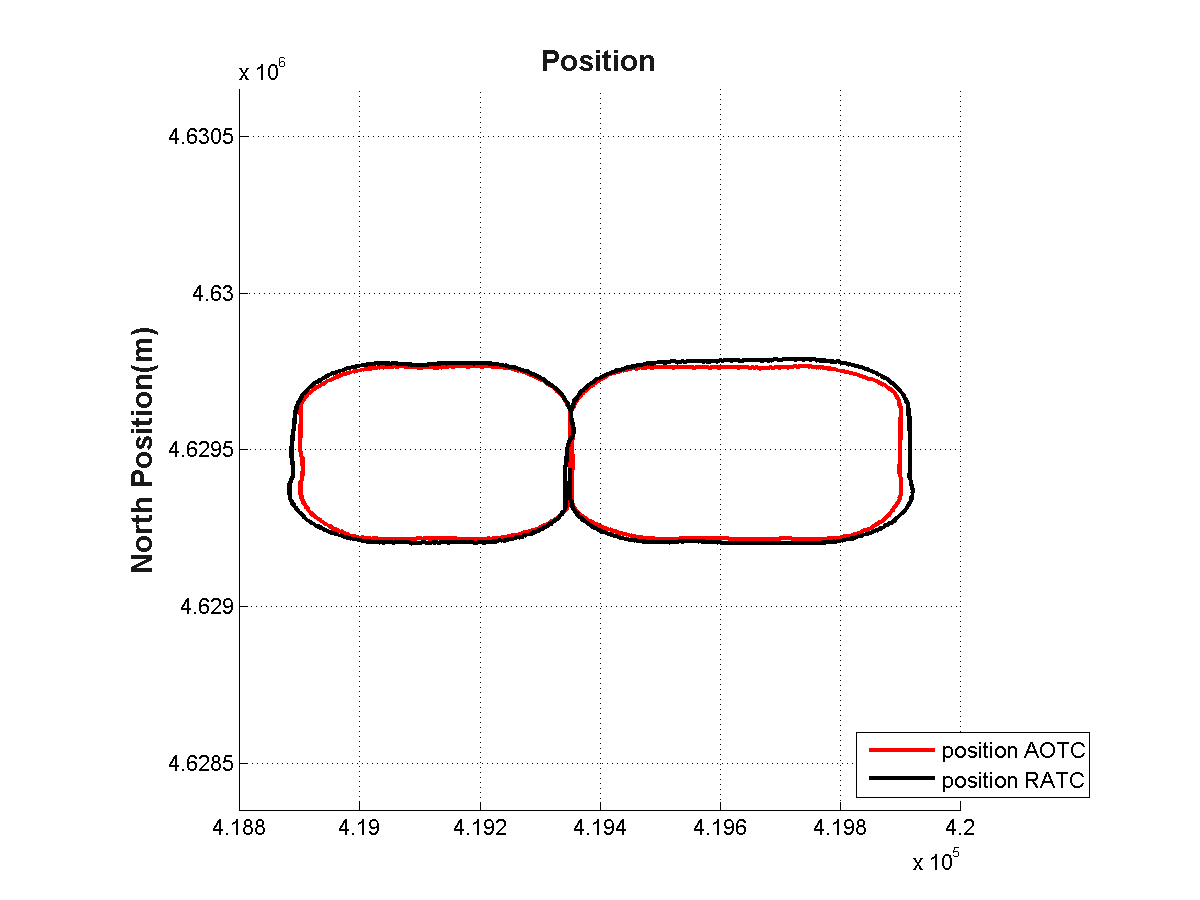}
	\caption{Position Flight \#1}
	\label{fig:FR_position}	
\end{figure}

In Figure \ref{fig:FR_position} changes between the trajectory schemes can be seen. Specifically, it is shown that the RATC had a harder time maintaining the desired flight line.  However, this was due to issues dealing with the abrupt change in desired course as noted in Section 8.6.   During this flight, the course gains were lowered in an attempt to counteract the effects of this problem. During later flights this issue was addressed and showed significant increase in controller error performance.   

\begin{figure}[H]
	\centering
	\includegraphics[width=8cm]{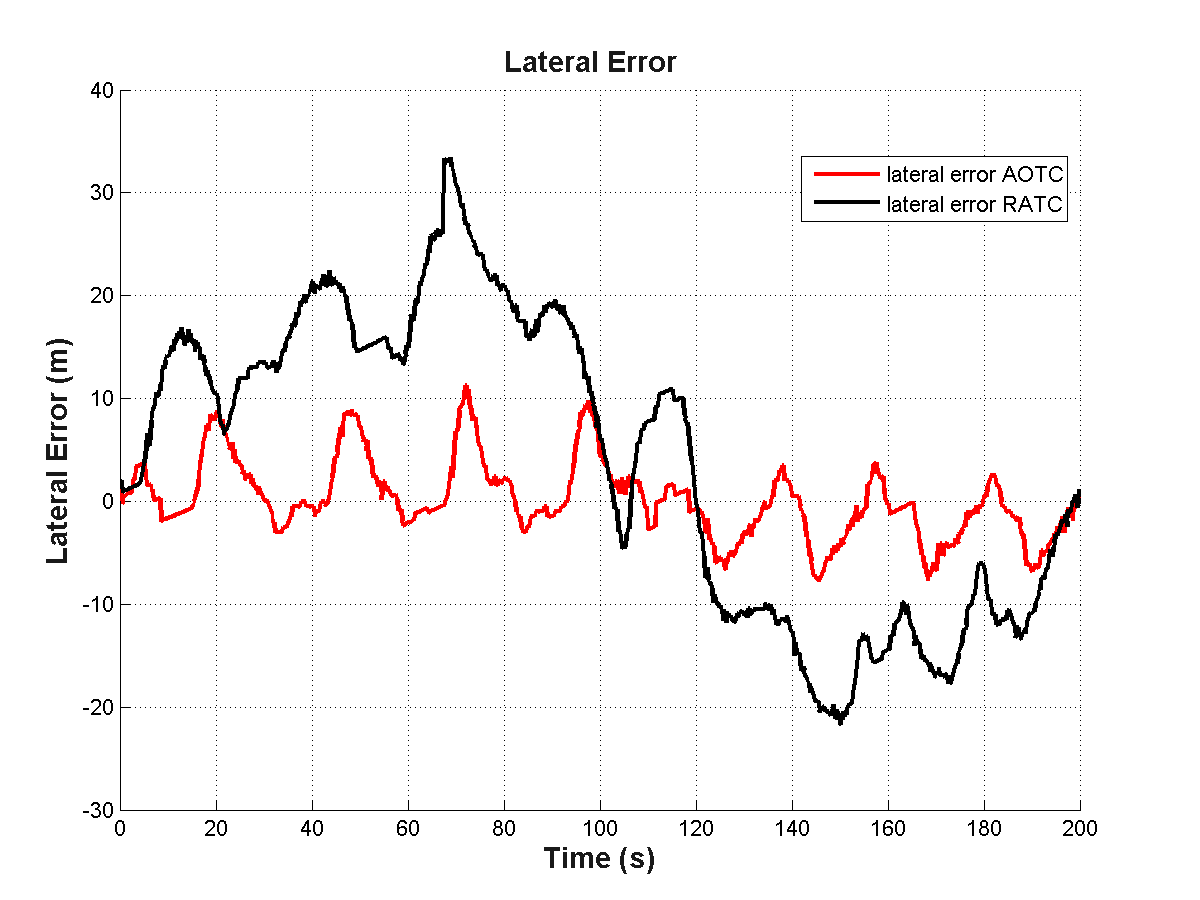}
	\caption{Lateral Error Flight \#1}
	\label{fig:FR_lat_err}
\end{figure}

\begin{figure}[H]
	\centering
	\includegraphics[width=8cm]{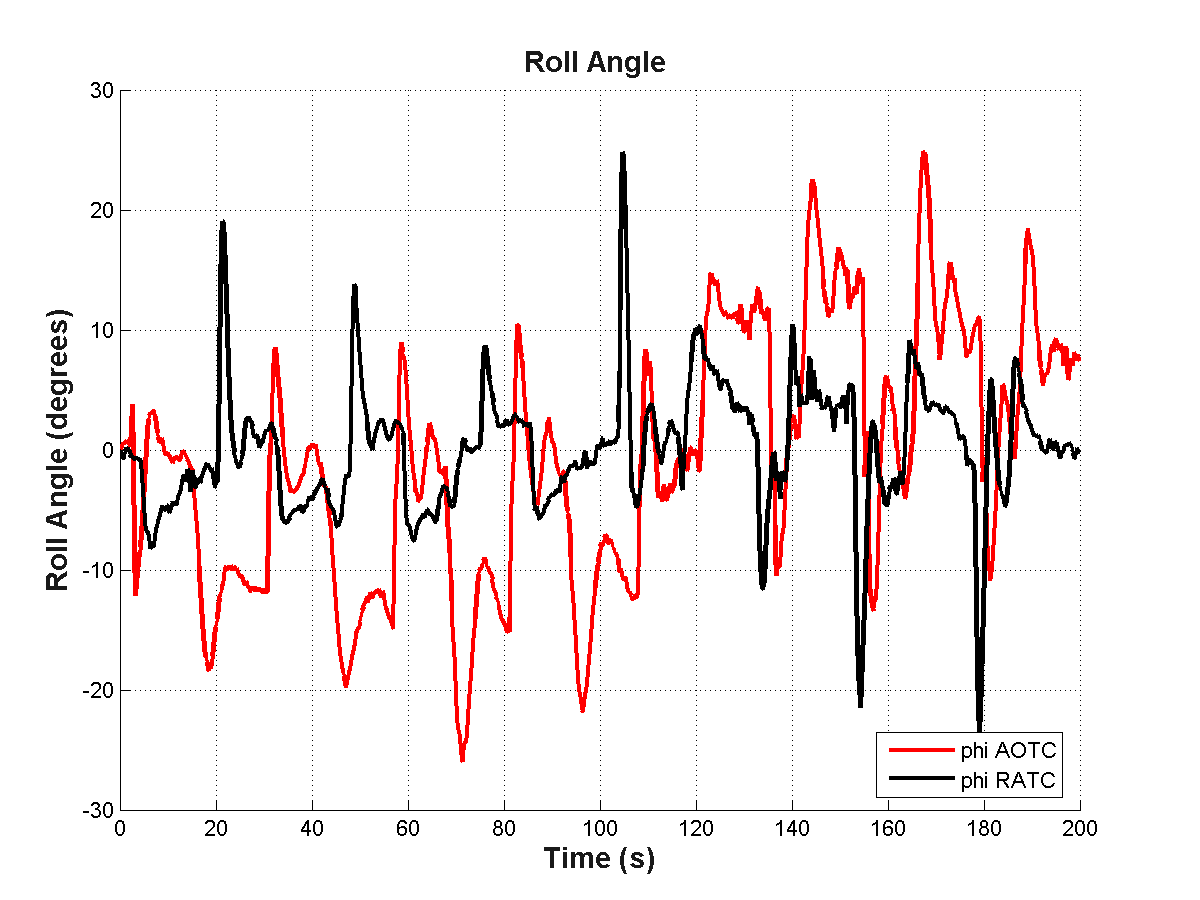}
	\caption{Roll Angle ($\phi$) Flight \#1}
	\label{fig:FR_phi}
\end{figure}

\begin{figure}[H]
	\centering
	\includegraphics[width=8cm]{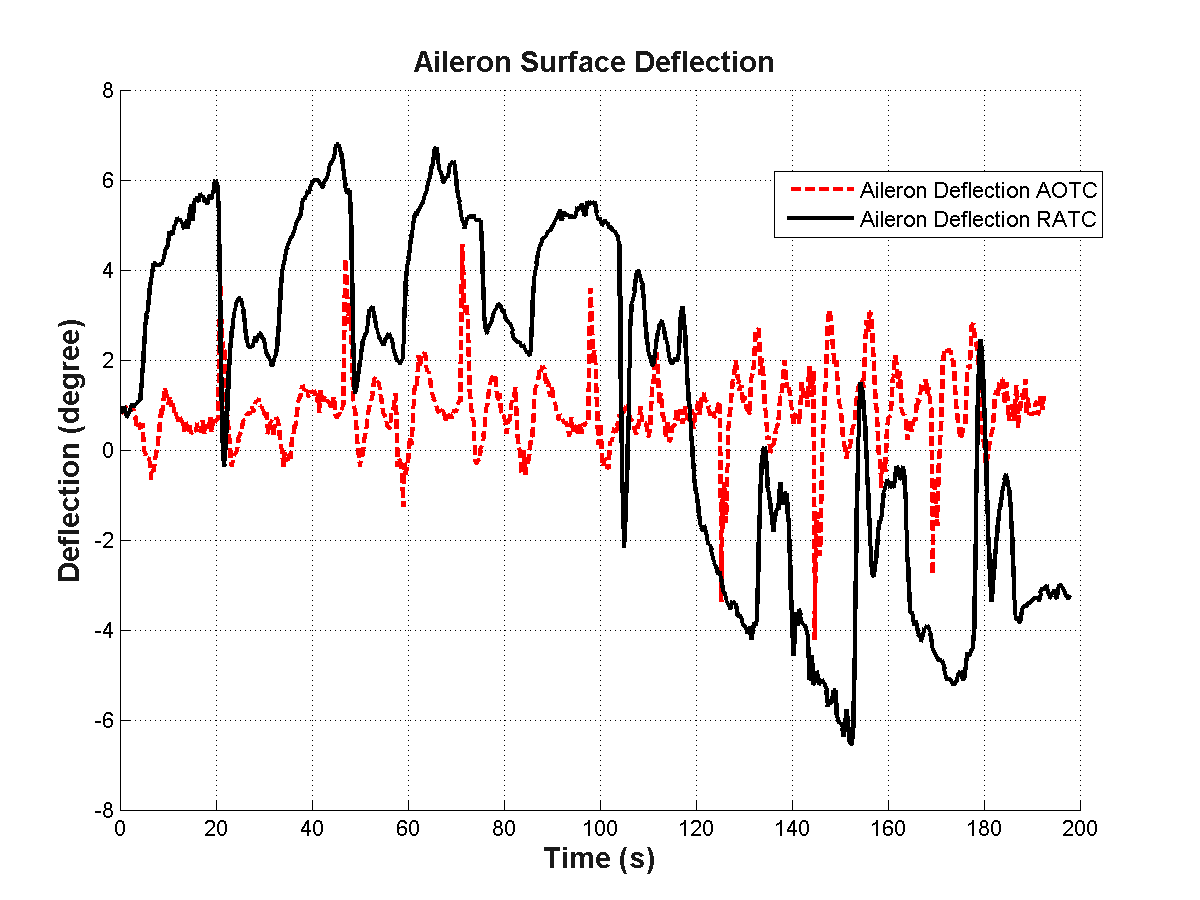}
	\caption{Aileron Control Surface Input Flight \#1}
	\label{fig:FR_CS_A}
\end{figure}

\begin{figure}[H]
	\centering
	\includegraphics[width=8cm]{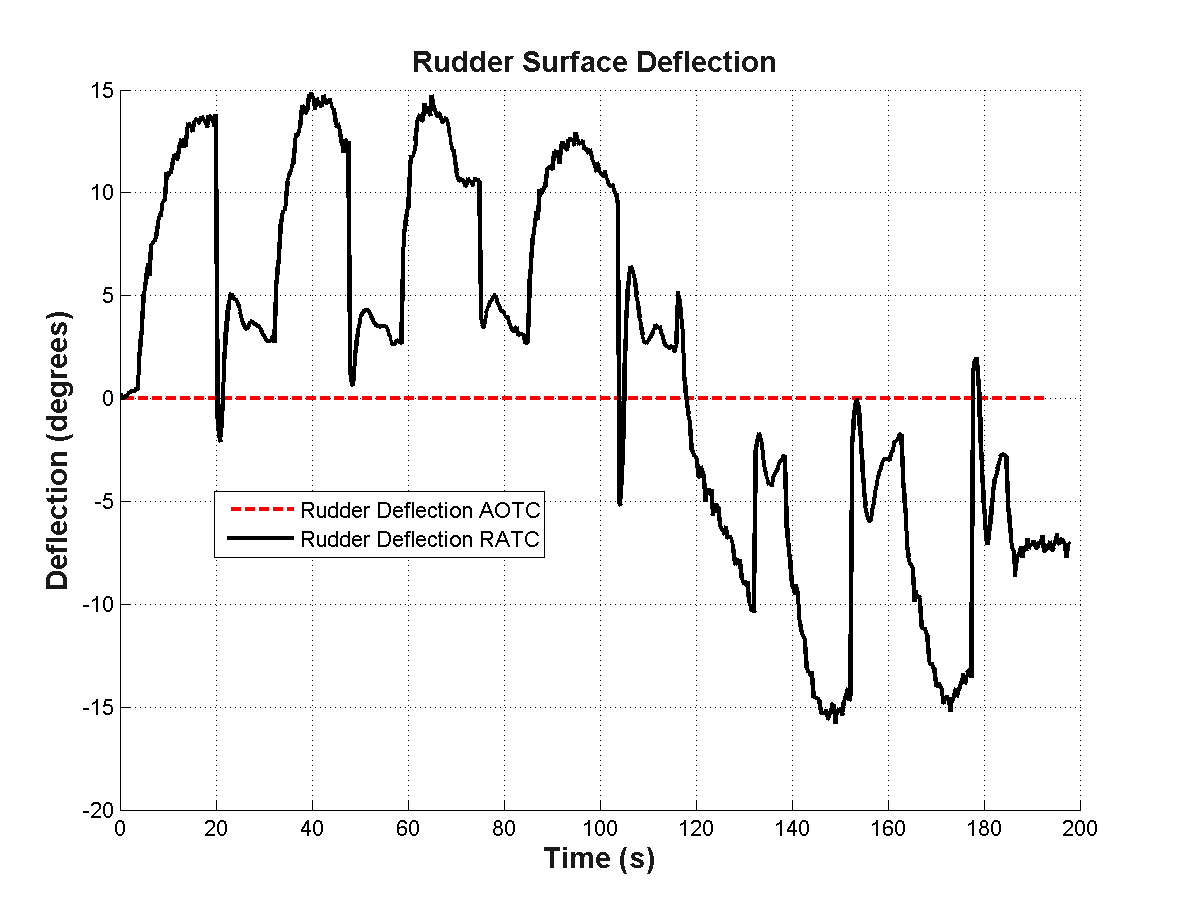}
	\caption{Rudder Control Surface Input Flight \#1}
	\label{fig:FR_CS_R}
\end{figure}

The total lateral error is shown in Figure \ref{fig:FR_lat_err}.   As was visible in \ref{fig:FR_position}, the RATC controller had a higher ground track error.  The RATC algorithm hit a maximum of 34 meters while the AOTC was only 12 meters.  While it seems rather high at nearly 200\% increase in image error, once the roll angle is factored into the error calculation, the lateral error is nearly an order of magnitude lower in contribution. 
Figure \ref{fig:FR_phi} shows the roll angle of the aircraft.   Notice that with the exception of the spikes (addressed in Section 8.6) the overall roll angle for RATC  stayed below 10 degrees.  However, AOTC averaged 10 degrees with initial correction angles of around 20 degrees.    
As was illustrated in the simulations, the control surface input (Figures \ref{fig:FR_CS_A} and \ref{fig:FR_CS_R}) was considerably higher in the RATC case.  However, even with the large disparity, the total movement for the rudder surface was below 16 degrees. 

\begin{figure}[H]
	\centering
	\includegraphics[width=8cm]{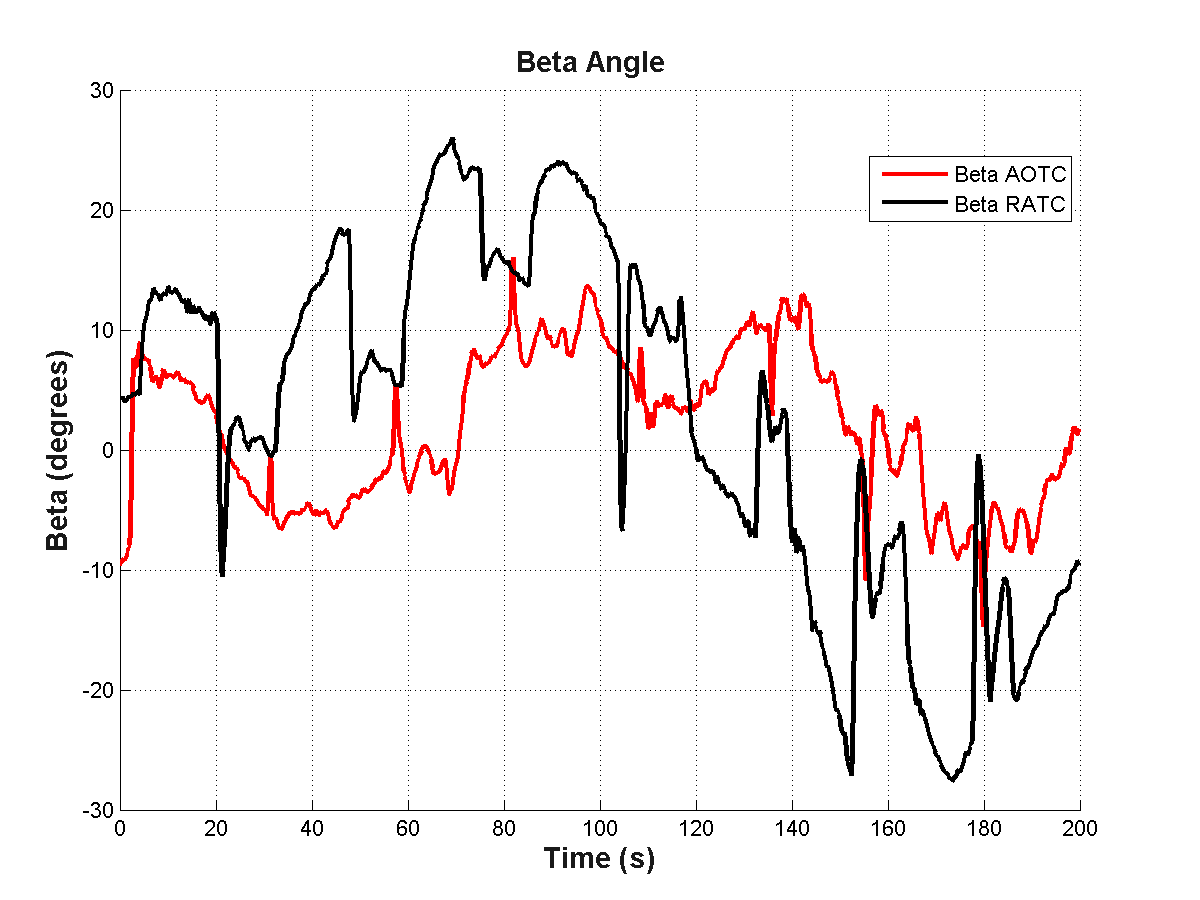}
	\caption{Beta  Flight \#1}
	\label{fig:FR_beta}
\end{figure}

Figure \ref{fig:FR_beta} shows the sideslip angle $\beta$ as a function of flight time.   During the flight tests, the sideslip was estimated via the IMU (VN200) and GPS heading data.     The inertial heading estimated from the VN200 was subtracted from the GPS course heading and the results were normalized to a $0-360 $ degree range.   Since light winds were present during the flight tests, the data presented is not 100\% accurate, but in the absence of a true sideslip 5 port Pitot tube sensor, they show the approximate comparisons.  The RATC algorithm had obviously higher sideslip angles.   This data trend was also verified with visual observations from the ground.

\begin{figure}[H]
	\centering
	\includegraphics[width=8cm]{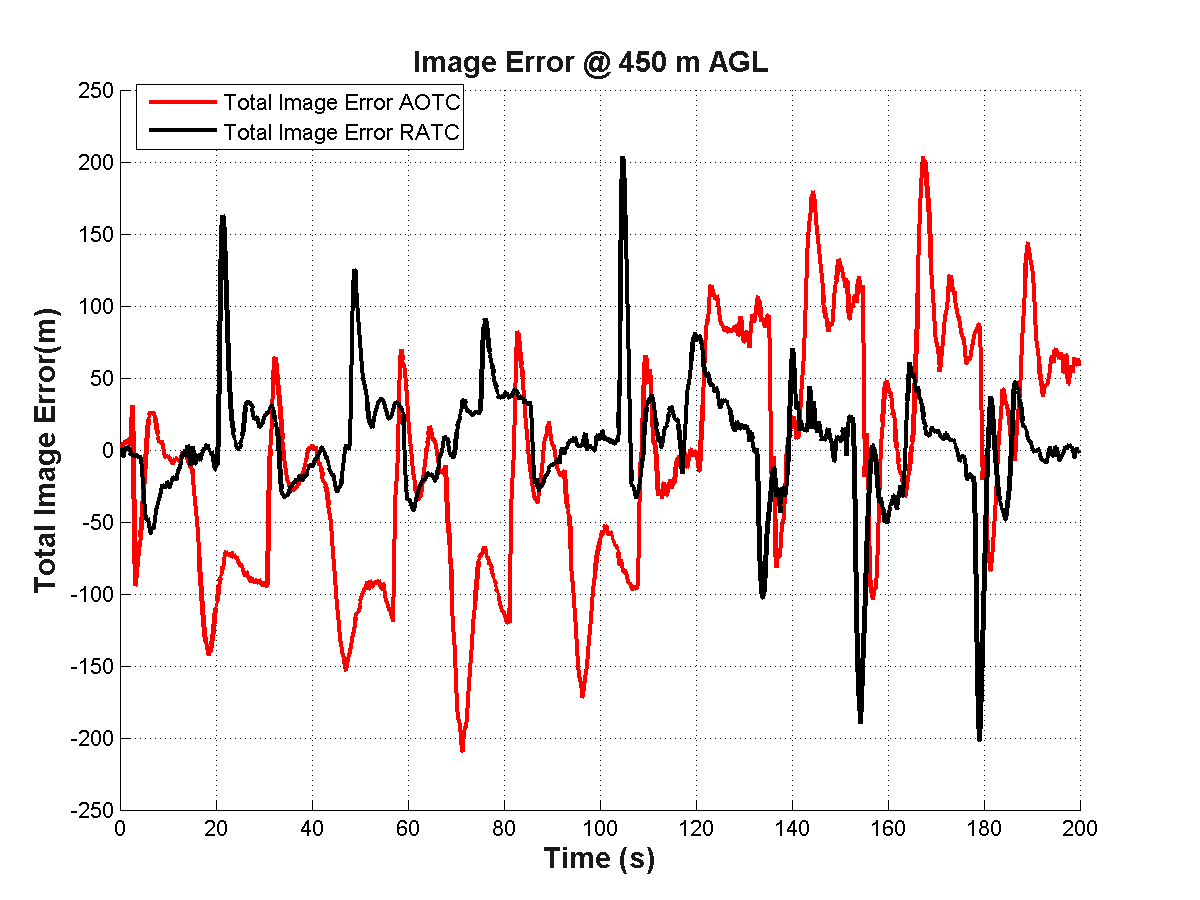}
	\caption{Total Image Error @450 meters AGL Flight \#1}
	\label{fig:FR_total_image_err}
\end{figure}

Figure \ref{fig:FR_total_image_err} shows the total image error vs flight time.   Even though the lateral error was 200\% higher in the RATC case, with the roll angle factored into the equation, RATC offers significant reductions in the total image error.

\subsection{Flight Plan \#2}

\begin{figure}[H]
	\centering
	\includegraphics[width=8cm]{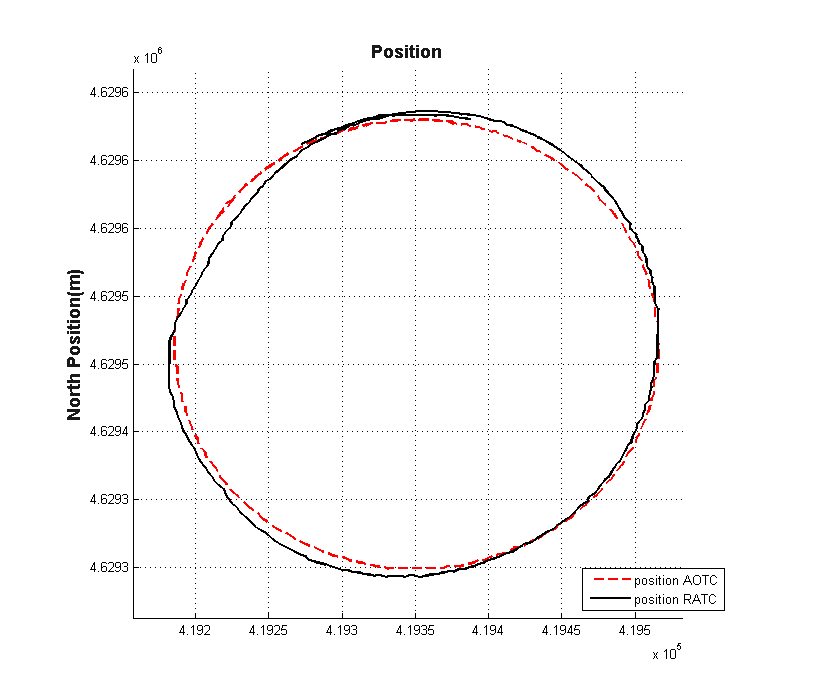}
	\caption{Position Flight \#2}
	\label{fig:FR_positionC}	
\end{figure}

Figure \ref{fig:FR_positionC} shows the circular pattern used in the energy calculations.   Visually, both algorithms perform well, with the RATC slightly more inconsistent in performance.   However, when compared to Figure  \ref{fig:FR_position}, a significant improvement in lateral error is noted.  An interesting observation from the maneuver can be seen in the mid left hand side of the RATC circle.     The estimated wind was nearly west to east and the large perturbation from nominal occurred as the aircraft pivoted into the wind while trying to maintain the desired trajectory.  Since banking the main wing can provide more side force than the thrust vector and lift from the vertical surface,  larger sideslip angles and rudder control inputs were necessary with RATC. As a result, the roll controller had a harder time keeping up with the rudder command and the presence of sideslip introduced trajectory tracking errors.     High sideslip angles are one of the principle drawbacks of the RATC method and are greatly influenced by the aerodynamic coefficients of the aircraft controlled.  

As noted in earlier, RATC has a larger lateral error with similar percentage in relative error when comparing to AOTC. However, with the issues of course angle discontinuities absent in this test, the gains were increased and the reduction in total lateral error was large when compared to Figure \ref{fig:FR_lat_err}.  
\begin{figure}[H]
	\centering
	\includegraphics[width=8cm]{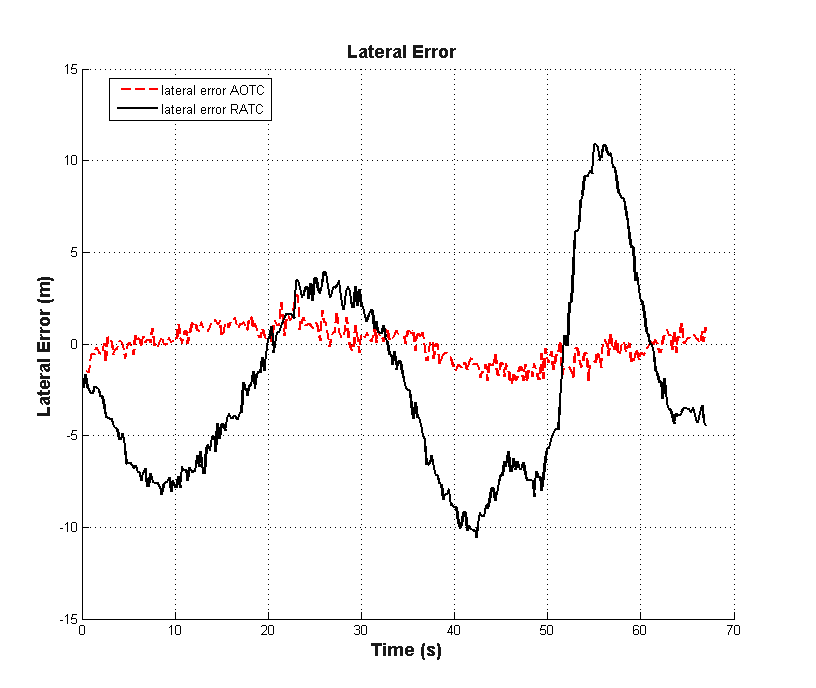}
	\caption{Lateral Error Flight \#2}
	\label{fig:FR_lat_errC}
\end{figure}

\begin{figure}[H]
	\centering
	\includegraphics[width=8cm]{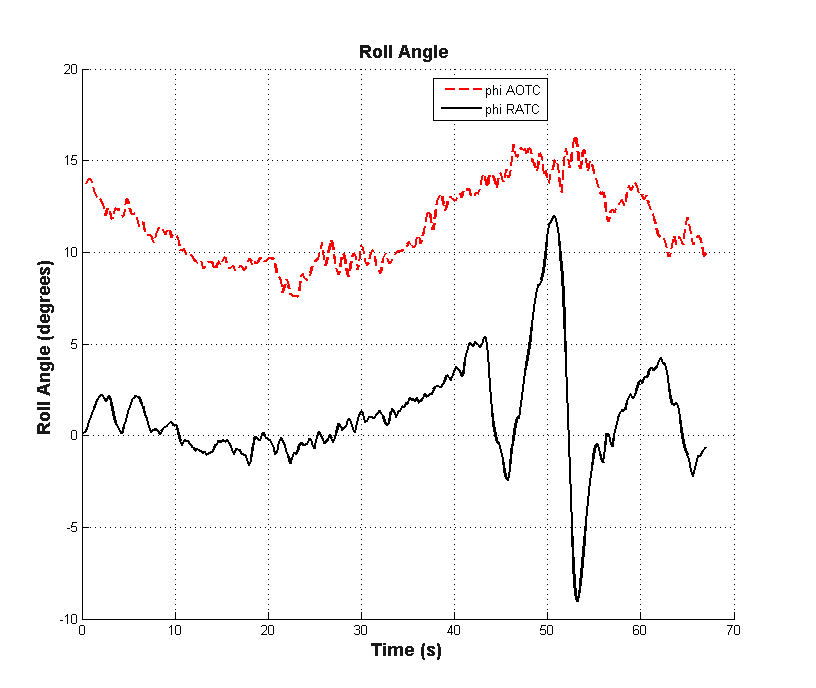}
	\caption{Roll Angle ($\phi$) Flight \#2}
	\label{fig:FR_phiC}
\end{figure}

\begin{figure}[H]
	\centering
	\includegraphics[width=8cm]{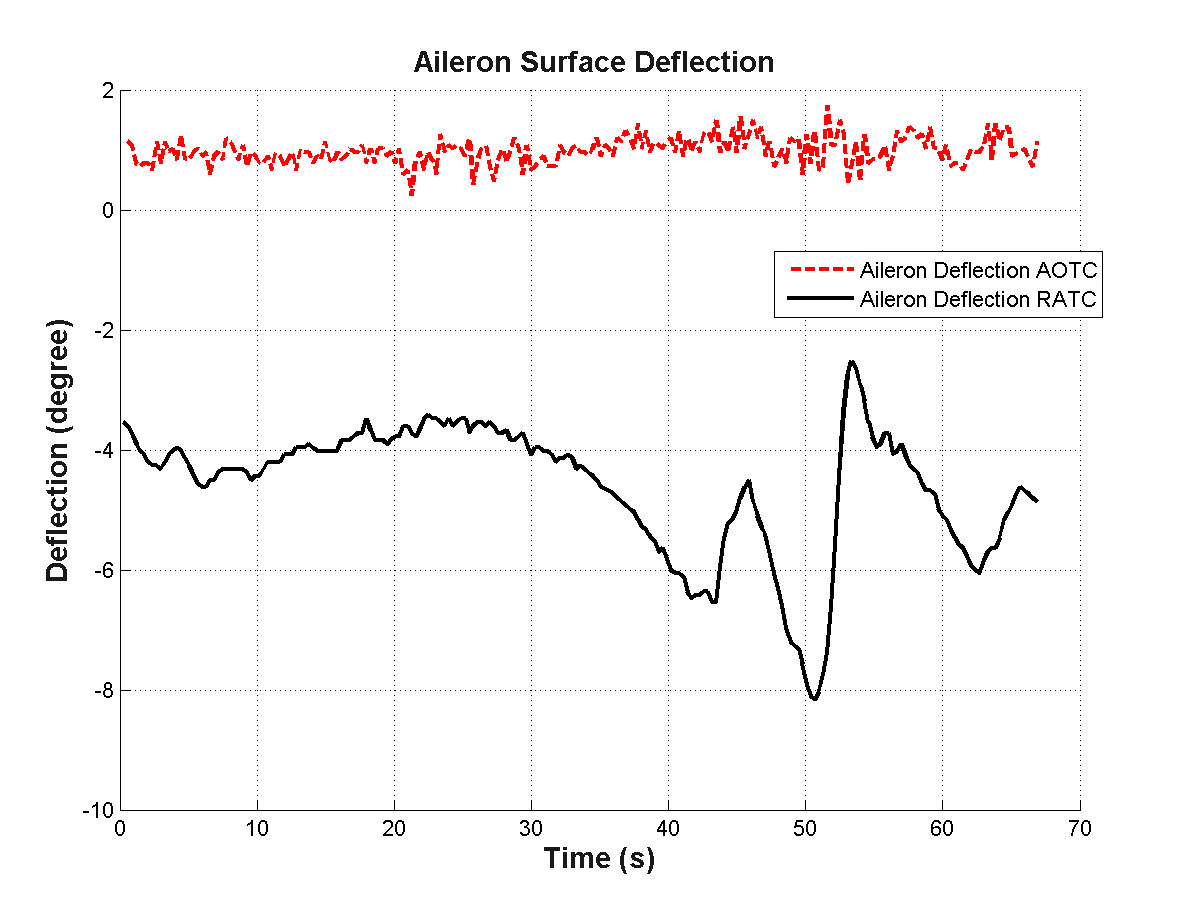}
	\caption{Aileron Control Surface Input Flight \#2}
	\label{fig:FR_CS_AC}
\end{figure}

Figure \ref{fig:FR_phiC} shows the roll angle of the aircraft.   The roll angle during RATC flight was considerably smaller than during AOTC flight, even when the aircraft was pivoting into the wind with large rudder control inputs.  
Figure \ref{fig:FR_CS_AC} shows the control surface deflection versus time.   Again, RATC requires significantly more control input than AOTC. Notice that in the RATC case, the aileron input follows the same trend as the rudder command.   This observation is noted in the future work section as a possibility for the incorporation of a feedforward aileron controller to give better response in roll control. 

\begin{figure}[H]
	\centering
	\includegraphics[width=8cm]{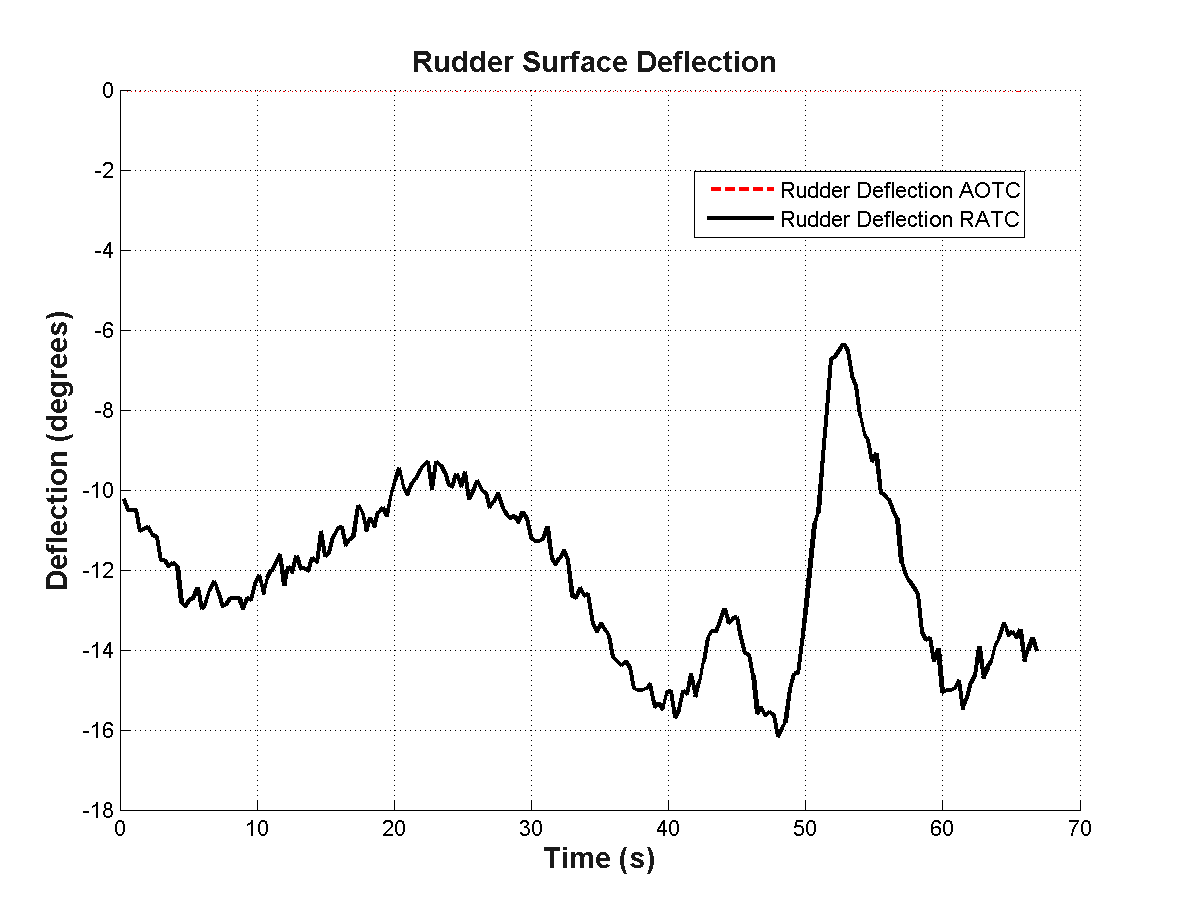}
	\caption{Rudder Control Surface Input Flight \#2}
	\label{fig:FR_CS_RC}
\end{figure}

\begin{figure}[H]
	\centering
	\includegraphics[width=8cm]{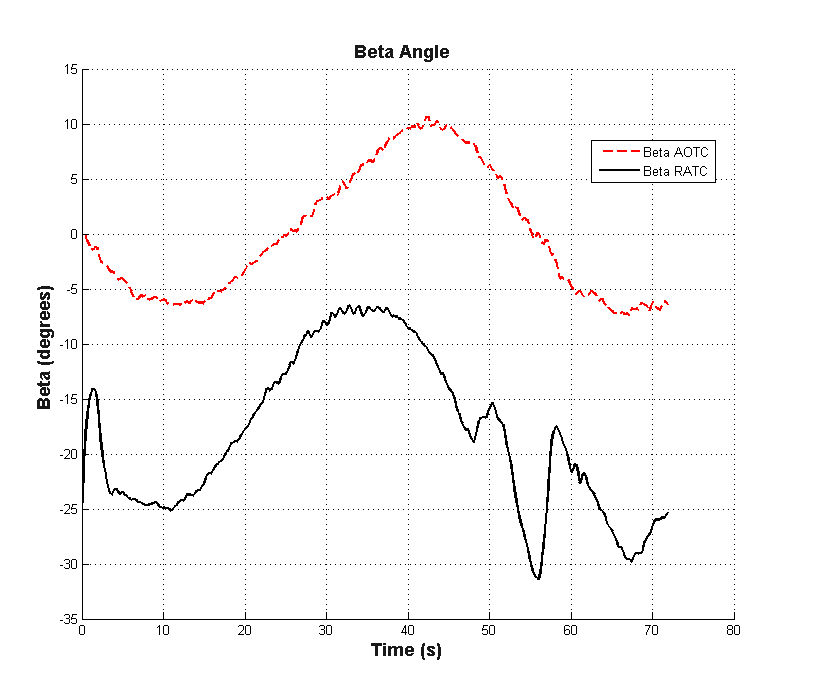}
	\caption{Beta Flight \#2}
	\label{fig:FR_betaC}
\end{figure}

As expected in Figure \ref{fig:FR_betaC}, the sideslip angles are higher in the RATC case.   Notice that as the aircraft pivots into the wind, the sideslip angle becomes very high, nearly 30  degrees.  To reduce the sideslip angle necessary to maintain trajectory control, physical changes to the airframe would be required to increase the aircraft's sideforce generated from $\beta$.

Even with the large disturbance in roll angle, the RATC image center stayed much closer to zero than the AOTC.   It is also noted that this is the most extreme maneuver that the RATC algorithm will  complete, since it was comprised of continuous turning flight both into and out of the wind vector.    

\begin{figure}[H]
	\centering
	\includegraphics[width=8cm]{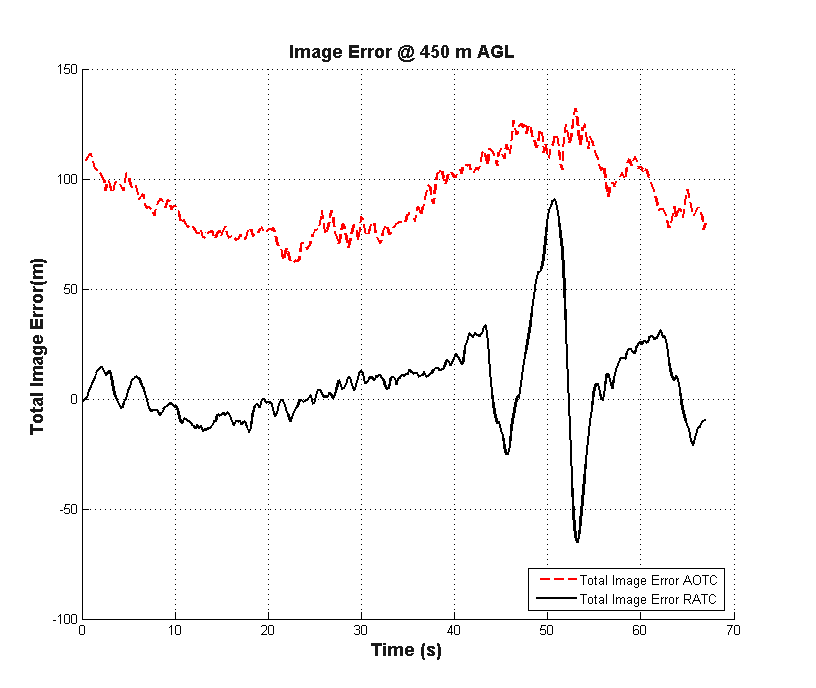}
	\caption{Total Image Error @450 meters AGL Flight \#2}
	\label{fig:FR_total_image_errC}
\end{figure}

\subsection{Flight Plan \#3}

The final flight included the use of the GoPro Hero2HD camera onboard for the payload verification in the thesis requirements.  This flight also incorporated a basic course slewing code to offset the issues discovered during the 10-28-2015 flights dealing with path planner desired course angle discontinuities.  As a result, the path and RATC gains were increased without the roll lag issues evident in Figure \ref{fig:FR_phi}.  In essence, the overall natural frequency of the RATC algorithm was increased without exceeding the natural frequency of the roll controller.  As shown in upcoming figures, this small change brought about large reductions in total image error  of the RATC algorithm, beyond those already presented.

\begin{figure}[h]
	\centering
	\includegraphics[width=8cm]{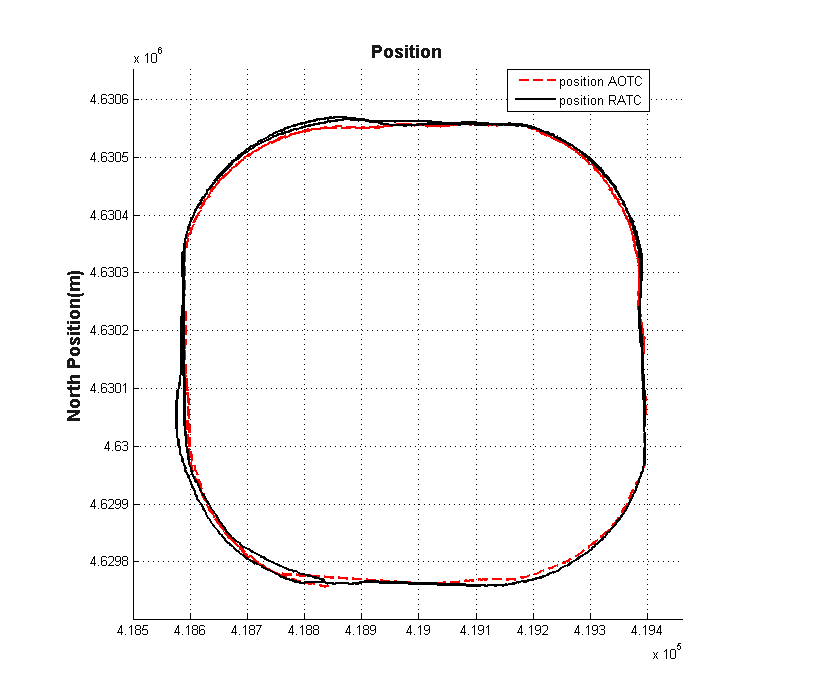}
	\caption{Position Flight \#3}
	\label{fig:FR_position1}	
\end{figure}
In comparison to Figure \ref{fig:FR_position}, the increase in RATC natural frequency  resulted in visibly lower  position error as seen in Figure \ref{fig:FR_position1}.   Both controllers in this case perform nearly equal, with the RATC algorithm having slight overshoot in the upwind (left hand side) corners.

\begin{figure}[H]
	\centering
	\includegraphics[width=8cm]{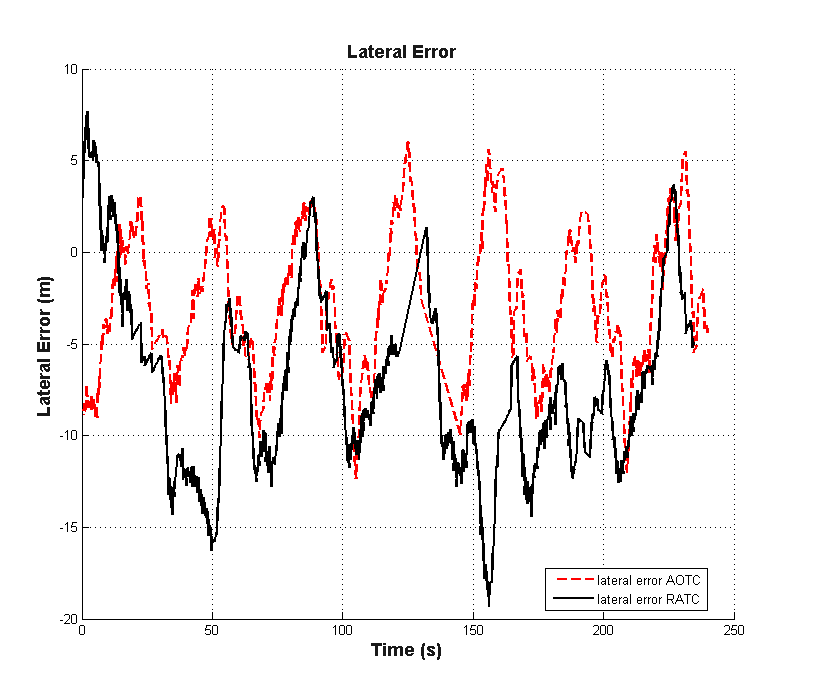}
	\caption{Lateral Error Flight \#3}
	\label{fig:FR_lat_err1}
\end{figure}

As inferred from  Figure \ref{fig:FR_position1}, the total lateral error as shown in  Figure \ref{fig:FR_lat_err1} was on the same order of magnitude for both methods.   With the exception of the data at 50, 150, and 200 seconds, the lateral error was almost identical.   When compared to Figure \ref{fig:FR_lat_err} where the disparity between methods was over 400\%, both RATC and AOTC performed at a similar level.   The overall shift of the error to a negative mean is a function of the trajectory controller derivations.   Both methods make assumptions about treating $\beta$ as an input disturbance.    Looking at Table \ref{tab:flight_results_lat_roll}, both methods see similar averages and variances.   For the AOTC case, the difference in both mean and standard deviation is less than 1.1 meters.  

\begin{figure}[H]
	\centering
	\includegraphics[width=8cm]{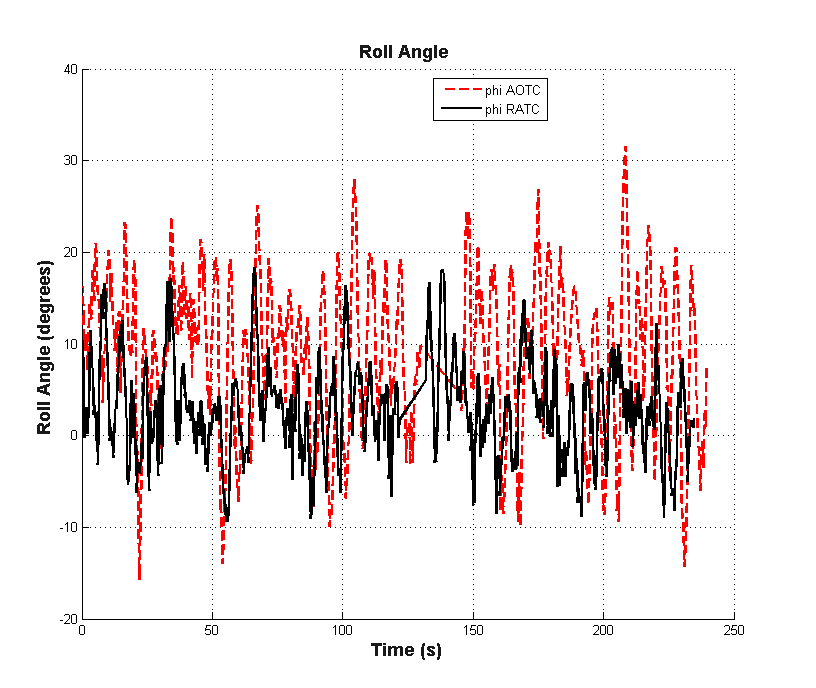}
	\caption{Roll Angle ($\phi$) Flight \#3}
	\label{fig:FR_phi1}
\end{figure}

As mentioned earlier, the alignment of the natural frequencies of the both the rudder and the roll angle controllers improved the overall performance of the algorithm.  Comparing the roll angle between Flight \#1 (Fig. \ref{fig:FR_phi})and Flight \#3 (Fig. \ref{fig:FR_phi1}), a marked difference is noted:  the large spikes in roll angle $\phi$ that occurred at each heading discontinuity  are no longer present.  As previously identified in both the simulation results and in the previous flight tests, the roll angle has a much larger bearing on the total error in the imagery than does the lateral cross-track error.    

\begin{figure}[H]
	\centering
	\includegraphics[width=8cm]{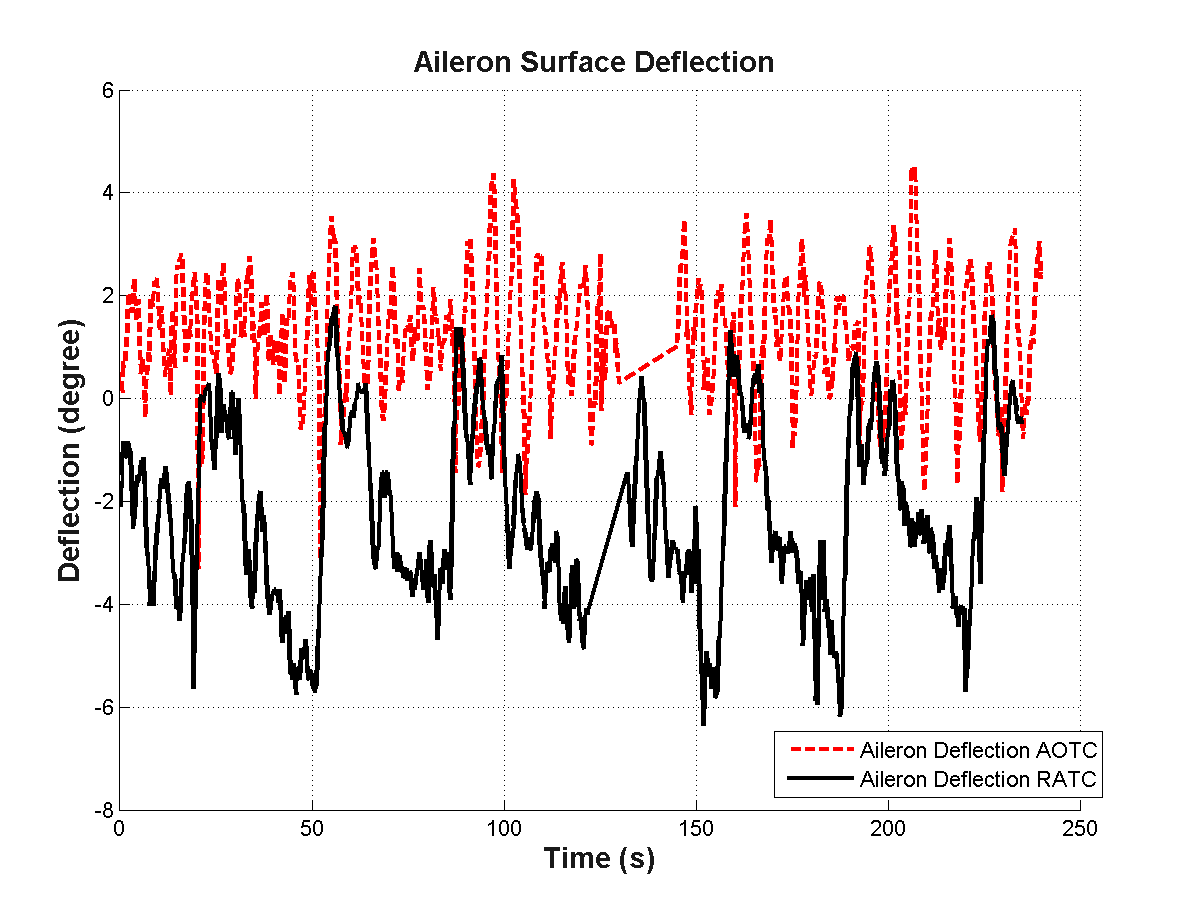}
	\caption{Aileron Control Surface Input Flight \#3}
	\label{fig:FR_CS_A1}
\end{figure}

\begin{figure}[H]
	\centering
	\includegraphics[width=8cm]{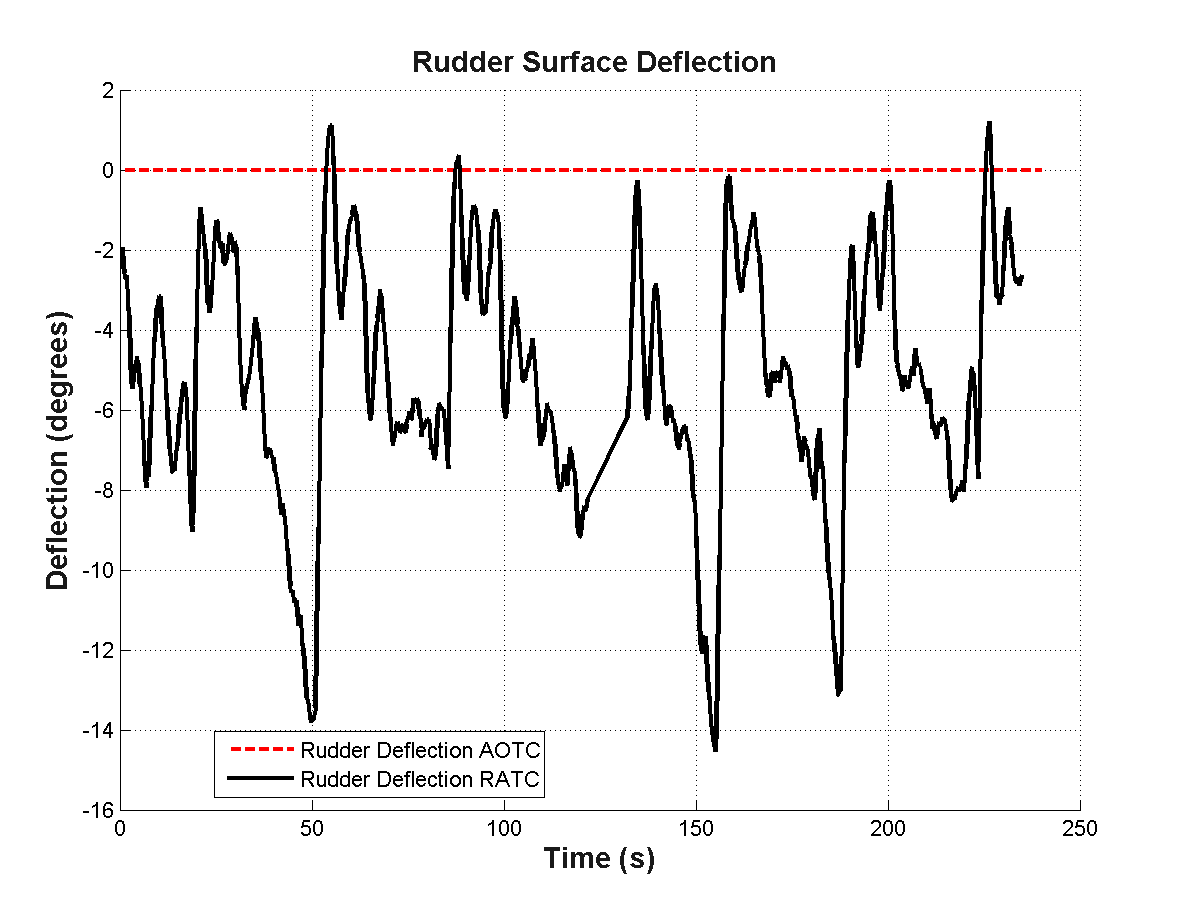}
	\caption{Rudder Control Surface Input Flight \#3}
	\label{fig:FR_CS_R1}
\end{figure}

Another interesting result of matching the natural frequencies of both the roll and yaw control loop, is a reduction in the sideslip angles observed during the tests.   As shown below in Table \ref{tab:flight_results_beta_comp}, there was a 40\% reduction in the mean and standard deviation of the sideslip between AOTC and RATC.   This is attributed to the roll controller being able to keep the aircraft level while still allowing increased rudder input.

\begin{table}[H]
	\centering
	\begin{tabular}{l|c}
		\hline
		\multirow{2}{*}{Flight Number} & Error Mean $\pm$ Standard Deviation (1-$\sigma$) \\ \hhline{~-}
		&  Sideslip Angle ($\beta$) \\ \hline
		
		\#1 - RATC ($2^{st}$ half) &  $-10.0 \pm 10.8 \si{\degree}$  \\ \hline			
		\#3 - RATC (GoPro) &   $-6.0 \pm 6.4 \si{\degree}$  \\  \hline
	\end{tabular}
	\caption{Flight Test Side Slip Angle Comparison}
	\label{tab:flight_results_beta_comp}
\end{table}

%	For flight \#3 mean = -6.0 +-6.4 degrees\\
%	for flight \#1 ($2^{nd}$ half) mean = -10.0 +- 10.8 degrees\\

\begin{figure}[H]
	\centering
	\includegraphics[width=8cm]{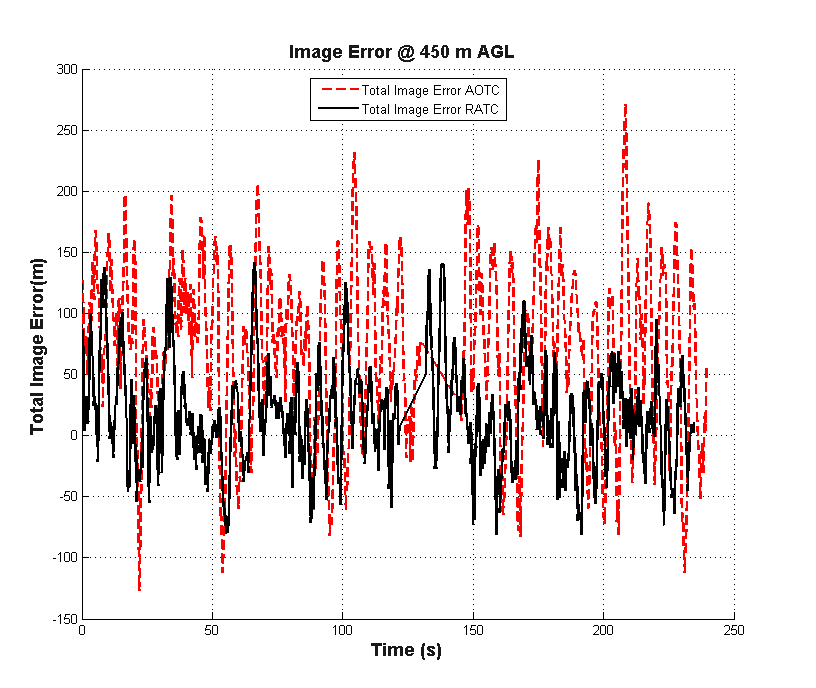}
	\caption{Total Image Error @450 meters AGL Flight \#3}
	\label{fig:FR_total_image_err1}
\end{figure}

\subsection{Payload Imagery Validation}
\begin{figure}[H]
	\centering
	\includegraphics[width=8cm]{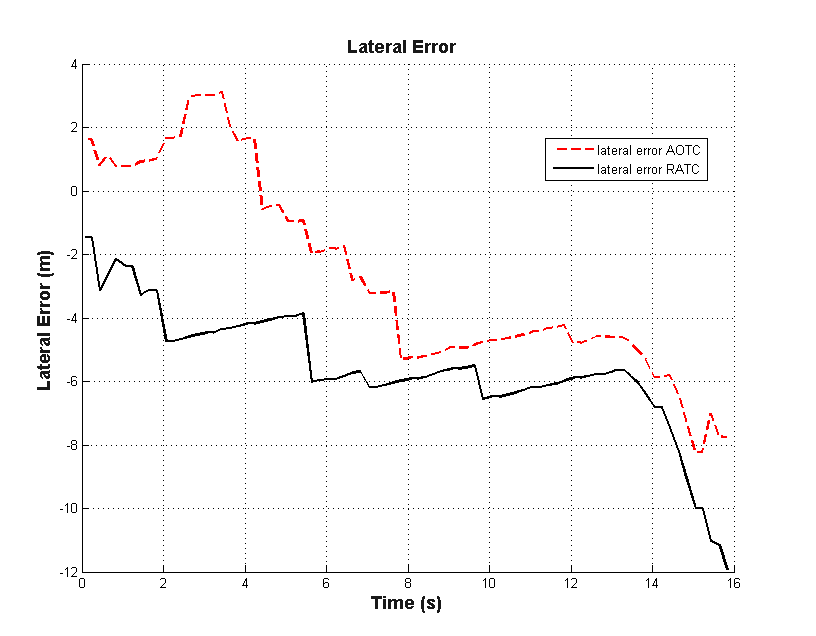}
	\caption{Lateral error in GoPro image dataset}
	\label{fig:LatErr_GP}
\end{figure}

\begin{figure}[H]
	\centering
	\includegraphics[width=8cm]{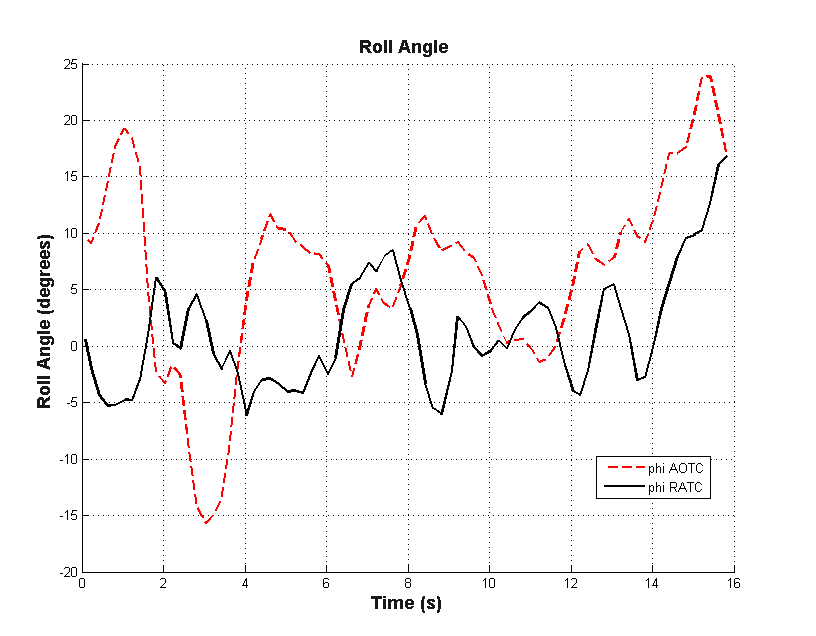}
	\caption{Roll angle in GoPro image dataset}
	\label{fig:RollAngle_GP}
\end{figure}

\begin{figure}[H]
	\centering
	\includegraphics[width=8cm]{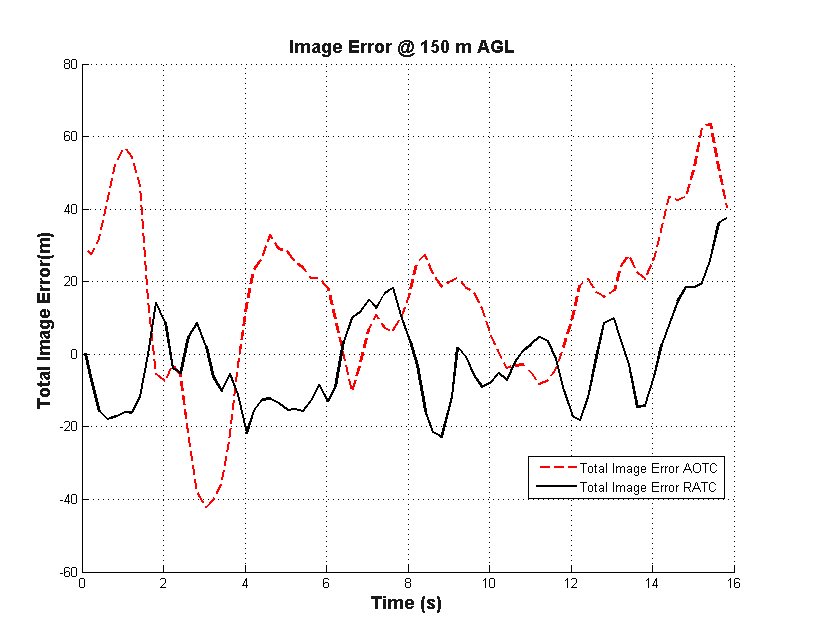}
	\caption{Image Error at flight altitude of 150m in GoPro dataset }
	\label{fig:imgErr_150_GP}
\end{figure}

To visualize what the previous graphs mean in terms of  data collection and potential,  Flight \#3 was flown over a series of roads and fence lines used as landmarks, confirming the IMU data shown in Figures \ref{fig:LatErr_GP} -\ref{fig:imgErr_150_GP}, as well as  Tables \ref{tab:flight_results} and \ref{tab:flight_results_lat_roll}.   Verification of data is difficult to see during the orbit/corner portions of the test due to faint and low contrast landmarks.   However,  the road and fence lines flown over during the straight flight line portions give good validation to the IMU data collected.  In addition, they provide valuable physical insight into the effect RATC has on payload imagery, more so than the raw IMU data shown throughout the rest of the paper.  Approximately 15 seconds of imagery is shown below, with an image sampled roughly every 3 seconds.  The wind vector present during the datasets is moving from the left to right (west to east) and was estimated at roughly 3 m/s. \\

\begin{figure}[H]
	\centering
	\includegraphics[width=10cm]{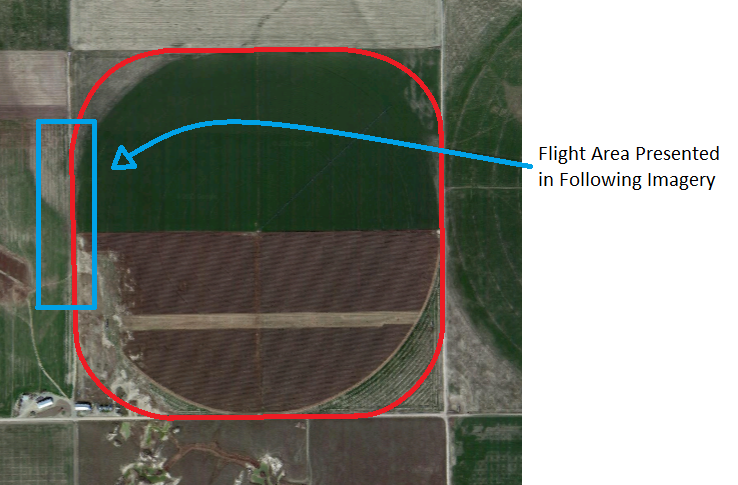}
	\caption{Flight Plan overlaid on Google Maps\textsuperscript{TM} }
	\label{fig:Flight_sample_area}
\end{figure}

At first glance, this small data set presented with the GoPro validation indicates a drop in performance of the RATC algorithm, compared to what has previously been shown.   Due to  low contrast in the payload imagery, the portion of flight selected for the comparison, was straight line - not orbit - following flight.   RATC offers the greatest and most notable benefits regarding lateral image error reduction during the orbiting or circular path portions of the flight plan.   However, RATC still had lower errors as shown in the following payload images.	

%	\begin{figure}[H]
%		\centering
%		\includegraphics[scale=0.7]{PICTURES_FLIGHT_RESULTS/aotc_line1}
%		\includegraphics[scale=0.7]{PICTURES_FLIGHT_RESULTS/aotc_line2}
%		\includegraphics[scale=0.7]{PICTURES_FLIGHT_RESULTS/aotc_line3}
%		\includegraphics[scale=0.7]{PICTURES_FLIGHT_RESULTS/aotc_line4}
%		\includegraphics[scale=0.7]{PICTURES_FLIGHT_RESULTS/aotc_line5}
%		\includegraphics[scale=0.7]{PICTURES_FLIGHT_RESULTS/aotc_line6}
%		
%		
%		\caption{Image set of AOTC payload performance at approximately 150m AGL}
%		\label{fig:aotc_gopro}
%	\end{figure}

%	\begin{figure}[H]
%		\centering
%		\includegraphics[scale=0.7]{PICTURES_FLIGHT_RESULTS/ratc_line1}
%		\includegraphics[scale=0.7]{PICTURES_FLIGHT_RESULTS/ratc_line2}
%		\includegraphics[scale=0.7]{PICTURES_FLIGHT_RESULTS/ratc_line3}
%		\includegraphics[scale=0.7]{PICTURES_FLIGHT_RESULTS/ratc_line4}
%		\includegraphics[scale=0.7]{PICTURES_FLIGHT_RESULTS/ratc_line5}
%		\includegraphics[scale=0.7]{PICTURES_FLIGHT_RESULTS/ratc_line6}
%		
%		
%		\caption{Image set of AOTC payload performance at approximately 150m AGL}
%		\label{fig:aotc_gopro}
%	\end{figure}


The first conclusion drawn about the differences between schemes is just how significant the roll angle error is in the total image error.   While that has been emphasized throughout both the simulations and the IMU data, seeing the visual difference adds depth to the benefits of using the RATC algorithm for real world payload data collection.  Looking at the first three images, the aircraft is coming out of the corner and attempting to follow the straight line portion of the flight plan.   The overall image stability is worse in the AOTC algorithm, AOTC having larger side-to-side swings  of the image center.   By Image Set 3 and 4 the AOTC algorithm had settled and relative errors between the frames and the methods were similar.   Image Set 5 starts the turn into the next corner.   The beginning of this turn shows some difference but it isn't until Image Set 6 that it is clear that the RATC method has better control of the roll angle in the corner turn.   This results in lower lateral image error and better image stability.

%During the majority AOTC image sets, a rotation about the z axis is noted,  resulting in a misalignment with the ground track. This is attributed to the aircraft wanting to naturally weather vane into the wind.    At the same time, RATC  stayed mostly aligned with the desired flight line.  Since accurate wind data was not available  using the sensors on the UAV it is hard to define all the specific changes that caused this phenomenon between methods to occur.   Both image sets were collected within 4 minutes of each other.   Figure \ref*{fig:LatErr_GP} gives some insight.   As the UAV was being pushed right (east) due to the wind, the RATC controller hit an equilibrium point with wind vector, gradient vector field and the rudder input.   In this case, RATC was applying enough rudder command to keep the %

%While wind variances can account for some of the difference, since both datasets were collected within a 4 minute window and accurate wind information at altitude could not be collected, it is difficult to say conclusively that the improvement of ground track error can be attributed completely to the RATC trying to maintain a $0 \si{\degree}$ (due North) heading angle.

\begin{table}[H]
	\centering
	\begin{tabular}{l|c|c|c}
		\hline
		\multirow{2}{*}{Case Number}   & \multicolumn{2}{c|}{True  Mean $\pm$ Standard Deviation (1-$\sigma$)} & {RMS Error} \\
		\hhline{~---}
		& Image Error 150m & Image Error 450m  & Image Error 450m \\
		\hline
		%	\multirow{2}{*}{Raaa (k)} & \multicolumn{2}{c|}{\multirow{2}{*}{this}} & 0.5 & 0.6\\
		%	\hhline{~~~--}            & \multicolumn{2}{c|}{}                      & 0.7 & 1.2 \\ \hline
		\rowcolor{Gray}
		\#1 - AOTC ($1^{st}$ half)  &  $-19.4 \pm 19.5 m$ &  $-58.5 \pm 60.6 m$  & $84.2m$  \\ 
		\rowcolor{Gray}
		\#1 - RATC ($1^{st}$ half) &  $13 \pm 13.7 m$    &  $9.1 \pm 38.2 m$&  $39.2m$\\ 
		\#1 - AOTC ($2^{st}$ half)  &  $18.5 \pm 20.7 m$    &  $56.3 \pm 62.6 m$  & $84.1m$ \\ 
		\#1 - RATC ($2^{st}$ half)  &  $-7.0 \pm 18.1 m$  &  $-1.5 \pm 47.6 m$ & $47.6m$ \\  
		\rowcolor{Gray} 
		\#2 - AOTC   &  $33.4 \pm 5.5 m$    &  $94.7 \pm 17.1 m$ &$96.2m$ \\   
		\rowcolor{Gray}
		\#2 - RATC    &  $0.9 \pm 8.0 m$    &  $6.7 \pm 22.4 m$ & $26.3m$ \\ 
		\#3 - AOTC   &  $21.7 \pm 23.1 m$    &  $67.2 \pm 68.3 m$ &  $95.7m$\\ 
		\#3 - RATC   &  $0.3 \pm 14.6 m$    &  $14.1 \pm 41.3 m$& $43.6m$ \\ 
		\rowcolor{Gray} 
		\#3 - AOTC (GoPro)   &  $15.6 \pm 22.9 m$    &  $49.9 \pm 68.4 m$ & $84.3m$  \\ 
		\rowcolor{Gray} 
		\#3 - RATC (GoPro)  &  $-1.9 \pm 13.2 m$    &  $4.6 \pm 39.7 m$& $39.7m$ \\\hline
		
	\end{tabular}
	
	\caption{Flight Test Total Error in Imagery Results}
	\label{tab:flight_results}
\end{table}

\begin{table}[H]
	\centering
	\begin{tabular}{l|c|c}
		\hline
		\multirow{2}{*}{Flight Number} & \multicolumn{2}{c}{True Error Mean $\pm$ Standard Deviation (1-$\sigma$)} \\ \hhline{~--}
		&  Lateral Error & Roll Angle \\
		\hline
		%	\multirow{2}{*}{Raaa (k)} & \multicolumn{2}{c|}{\multirow{2}{*}{this}} & 0.5 & 0.6\\
		%	\hhline{~~~--}            & \multicolumn{2}{c|}{}                      & 0.7 & 1.2 \\ \hline
		\rowcolor{Gray}
		\#1 - AOTC ($1^{st}$ half) & $2.0 \pm 3.5 m$   &   $-7.3 \pm 7.6 \si{\degree}$  \\ 
		\rowcolor{Gray}
		\#1 - RATC ($1^{st}$ half) & $15.1 \pm 7.5 m$   & $-0.8 \pm 4.6 \si{\degree}$  \\ 
		\#1 - AOTC ($2^{st}$ half) & $-2.1 \pm 2.8 m$   & $7.1 \pm 7.8 \si{\degree}$  \\ 
		\#1 - RATC ($2^{st}$ half) & $-10.0 \pm 8.0 m$  &   $1.1 \pm 5.7 \si{\degree}$  \\ 
		\rowcolor{Gray} 
		\#2 - AOTC  &  $-0.04 \pm 0.9 m$   &   $11.5 \pm 2.1 \si{\degree}$  \\ 
		\rowcolor{Gray}
		\#2 - RATC  &    $-2.3 \pm 5.2 m$  &   $1.1 \pm 2.9 \si{\degree}$  \\ 
		\#3 - AOTC &  $-2.7\pm 3.9 m$   &   $8.5 \pm 8.4 \si{\degree}$  \\ 
		\#3 - RATC &  $-7.2 \pm 5.0 m$   &   $2.6 \pm 5.1 \si{\degree}$  \\ 
		\rowcolor{Gray} 
		\#3 - AOTC (GoPro) &  $-2.7\pm 3.2 m$   &   $6.4 \pm 8.4 \si{\degree}$  \\ 
		\rowcolor{Gray} 
		\#3 - RATC (GoPro)&  $-5.6 \pm 2.0 m$   &   $1.2\pm 5.0 \si{\degree}$  \\  \hline
		
	\end{tabular}
	\caption{Flight Test Lateral and Roll Angle Error Results}
	\label{tab:flight_results_lat_roll}
\end{table}

\section{Special Considerations and Implications on Path Planning Algorithms}\label{s: Flight Paths}
\begin{figure}[h]
	\centering
	%\ContinuedFloat
	%	\subfloat[]{%
	\includegraphics[width=8cm]{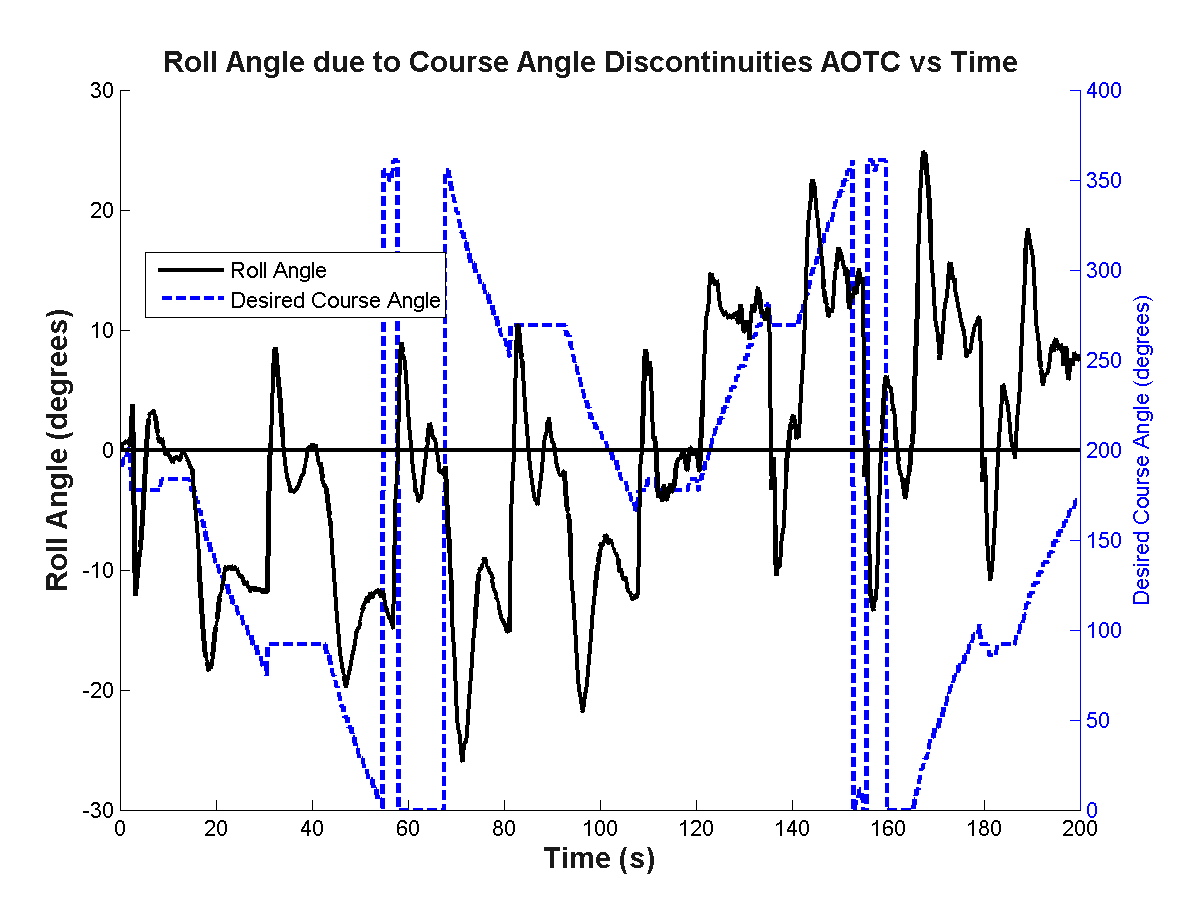}
	\caption{Desired Course Angle Discontinuity AOTC}
	\label{fig:course_discont_AOTC}
\end{figure}

\begin{figure}[h]
	\centering
	%\ContinuedFloat
	%	\subfloat[]{%
	\includegraphics[width=8cm]{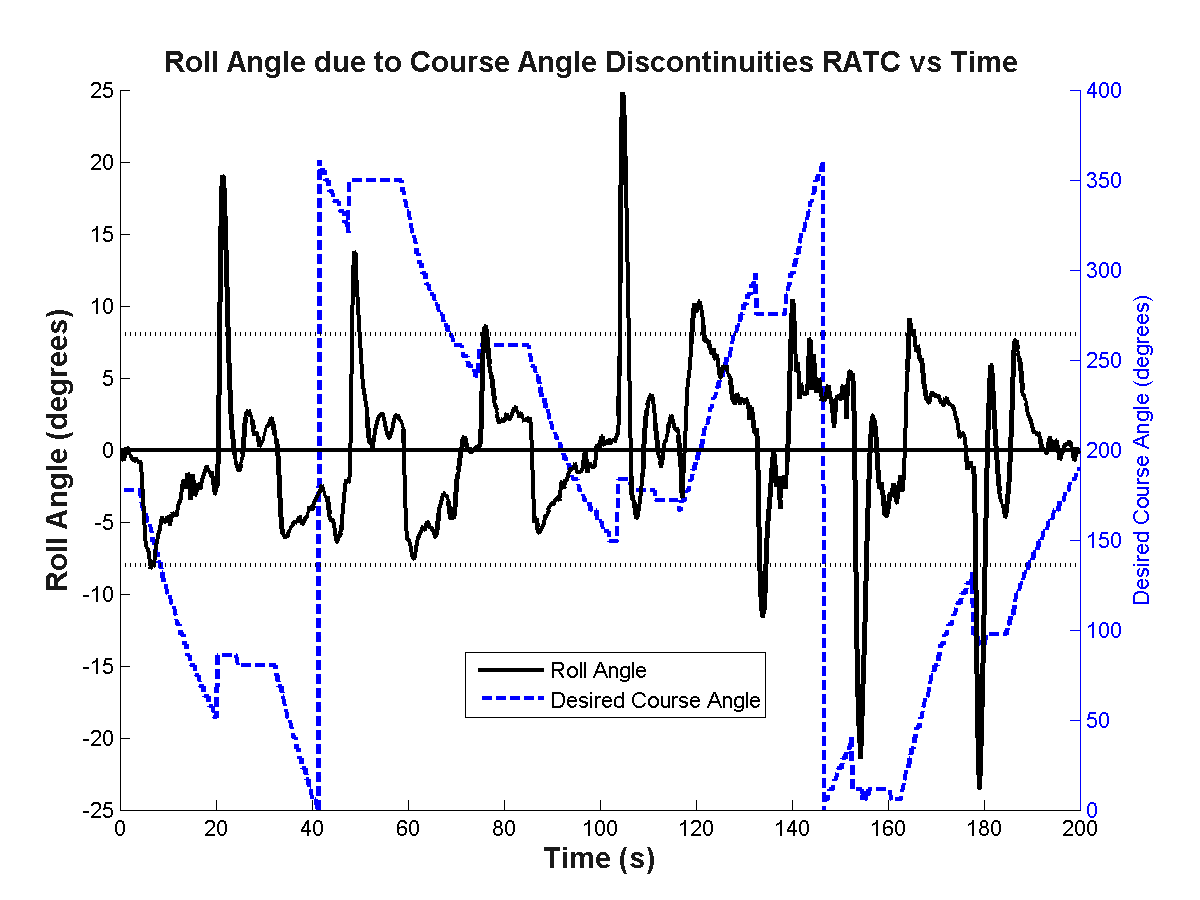}
	\caption{Desired Course Angle Discontinuity RATC}
	\label{fig:course_discont_RATC}
\end{figure}

\begin{figure}[H]
	\centering
	%\ContinuedFloat
	%	\subfloat[]{%
	\includegraphics[width=8cm]{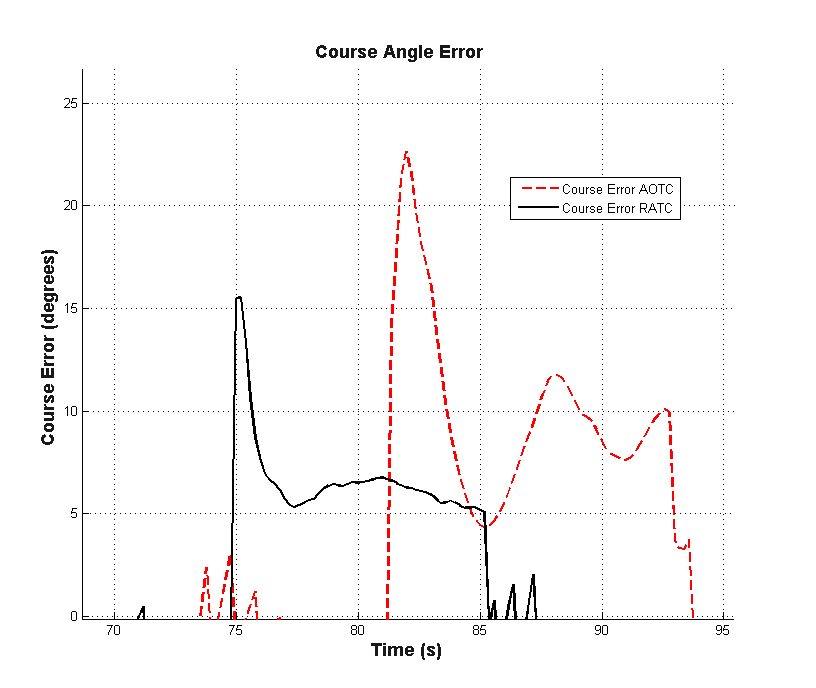}
	\caption{Course Error Response Uncorrected}
	\label{fig:course_response_fast}
\end{figure}

\begin{figure}[H]
	\centering
	%\ContinuedFloat
	%	\subfloat[]{%
	\includegraphics[width=8cm]{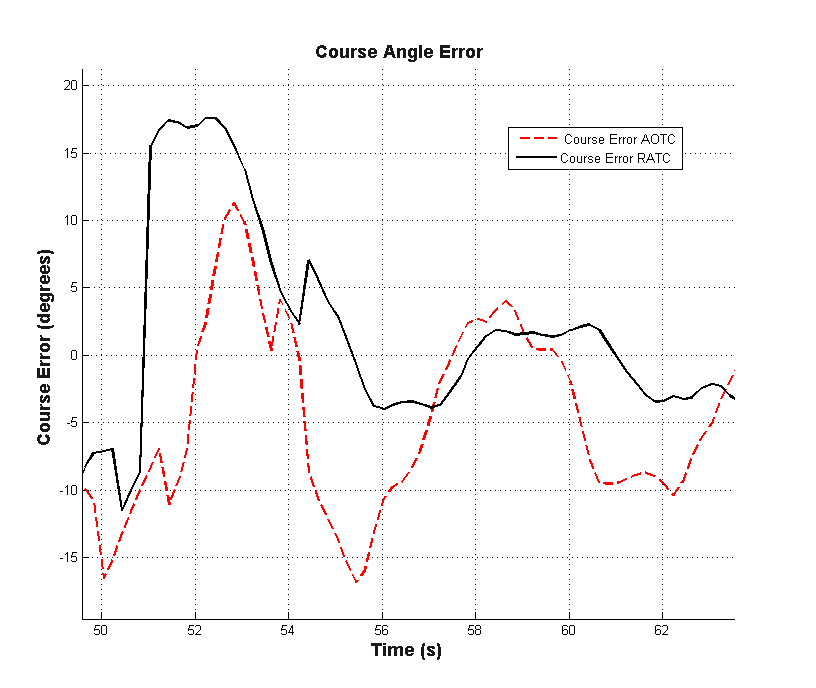}
	\caption{Course Error Response Corrected}
	\label{fig:course_response_slewed}
\end{figure}

Due to the rudder control surface being able to command heading changes faster than AOTC \cite{Ahsan2012}, special considerations must be taken when using AOTC-based path managers/path planners.  It was discovered that the implementation of the aggressive natural frequencies and corresponding control gains found in the simulation model were not realistic.  The Aerosonde simulation  had both an aggressive natural frequency for the yaw correction loop and an even more aggressive natural frequency for the roll correction loop.   When attempts were made to replicate this in hardware, severe oscillations were discovered in the roll controller.   As a result, the natural frequency was lowered in the roll loop.   To keep good lateral error performance, the RATC yaw natural frequency was kept high instead of being lowered like the roll loop (see figure \ref{fig:course_response_fast}).    This compromise seemed to work well.  However, as illustrated in Figure \ref{fig:FR_phi}, large spikes in the roll angle were evident after orbit following flight regimes.   Extensive analysis was performed and it was discovered that the root cause of the problem was discontinuity in the heading command.   While the roll correction controller was normally able to keep up with the high gains of the yaw loop, the roll controller was overwhelmed when these discontinuities were present.  While this phenomenon also existed with AOTC,  it didn't effect control stability and therefore was never addressed.  As shown in Figures \ref{fig:course_discont_AOTC} and \ref{fig:course_discont_RATC}, at the point of discontinuity, the AOTC method rolls from a high roll angle to a significantly lower but opposite bank angle.   Contrast this with the RATC case, where at the point of discontinuity, it jumps from a very low bank angle to the large spikes described above.     As a basic corrective measure, the course command was restricted to a specified rate of change.   If the course command step change went above this rate of change, the command was artificially slewed and attenuated  before being sent to the yaw control loop.  This change was very effective at choking the higher natural frequencies in the yaw control loop and allowed the roll correction loop to maintain control.   The choking of the high natural frequencies can be seen in figure \ref{fig:course_response_slewed}.


In figure \ref{fig:course_discont_AOTC}, numerous discontinuities can be seen, for example at the 30, 80, and 180 second time stamps. At each time stamp, an opposite but overall decreased roll angle is observed.   It can also be seen that the response is fairly smooth, indicating that the lower natural frequency of the roll controller for AOTC flight doesn't destabilize the system with the discontinuity.

Unlike figure \ref{fig:course_discont_AOTC}, the RATC response is significantly different than AOTC case even with a  similar heading change (see Figure \ref{fig:course_discont_RATC}).   Notice that the roll angle goes from controlled state to a sharp impulse as the discontinuity enters the yaw controller.    As noted earlier, the faster natural frequency of the yaw loop reacted quickly to the command change but the roll controller could not keep up.    


Notice the difference in settling times between the two methods as shown in Figure \ref{fig:course_response_fast}.  As described above, the RATC yaw response natural frequency was left high to decrease the overall lateral error.  The natural frequency of the RATC yaw controller was nearly double that of the roll controller.    This scenario didn't cause problems in normal flight, but was only evident when there were discontinuities in the course angle resulting in a step change in the yaw error.

Figure \ref{fig:course_response_slewed} shows the difference in the response time of the yaw loop when discontinuities in the desired course angle were slewed.  Unlike \ref{fig:course_response_fast}, where the response of the yaw loop was nearly twice that of the roll controller, with the active attenuation of the discontinuity errors, the RATC yaw bandwidth was slightly slower than that of the roll control loop.   This allowed the roll control loop to keep pace with the yaw loop during the step change, yet did not effect its normal error correction thus resulting in lower lateral errors.   Both of these effects are manifest when comparing Figure \ref{fig:FR_phi} to \ref{fig:FR_phi1} with the later graph showing a considerable improvement in roll performance between the two RATC implementations.  
\section{Concusion}\label{s: Conclusions}
	 A rudder augmented trajectory correction method for small unmanned aerial vehicles is discussed in this paper.   The goal of this type of controller is to minimize the lateral image errors of body fixed non-gimbaled  cameras.   We present a comparison to current aileron only trajectory correction autopilots.   Simulation and flight test results are presented that show significant reduction in the roll angle present during trajectory correction resulting in a large effect on total  flight line image deviations.
	\section*{Acknowledgment}
	The authors would like to acknowledge Dr. Mac McKee of the Utah Water
	Research Laboratory. This work is supported by Utah Water Research
	Laboratory MLF 2006-2015. 
	
	%----------------------------------------------------------------------------------------
	%	BIBLOGRAPHY
	
	% produces the bibliography section when processed by BibTeX
	\bibliographystyle{aiaa}
	\bibliography{ref}

\end{document}